\documentclass[epj]{svjour}
\usepackage{graphics}
\usepackage{amsmath}
\usepackage{amsfonts}
\usepackage{amssymb}

\begin{document}
\title{Dispersion and entropy-like measures of multidimensional harmonic systems. Application to Rydberg states and high-dimensional oscillators}
\author{J. S. Dehesa\inst{1} \and I. V. Toranzo\inst{2}
\thanks{\emph{Present address:} dehesa@ugr.es}%
}                     
\offprints{}          
\institute{Instituto Carlos I de F\'{\i}sica Te\'orica y Computacional, Universidad de Granada, 18071 Granada, Spain and Departamento de F\'{\i}sica At\'{o}mica, Molecular y Nuclear, Universidad de Granada, 18071 Granada, Spain \and Departamento de Matem\'atica Aplicada, Universidad Rey Juan Carlos, 28933 Madrid, Spain}
\date{Received: date / Revised version: date}
%
\abstract{
The spreading properties of the stationary states of the quantum multidimensional harmonic oscillator are analytically discussed by means of the main dispersion measures (radial expectation values) and the fundamental entropy-like quantities (Fisher information, Shannon and R\'enyi entropies, disequilibrium) of its quantum probability distribution together with their associated uncertainty relations. They are explicitly given, at times in a closed compact form, by means of  the potential parameters (oscillator strength, dimensionality, $D$) and the hyperquantum numbers $(n_r,\mu_1,\mu_2,\ldots,\mu_{D-1})$ which characterize the state. Emphasis is placed on the highly-excited Rydberg (high radial hyperquantum number $n_r$, fixed $D$) and the high-dimensional (high $D$, fixed hyperquantum numbers) states. We have used a methodology where the theoretical determination of the integral functionals of the Laguerre and Gegenbauer polynomials, which describe the spreading quantities, leans heavily on the algebraic properties and asymptotical behavior of some weighted $\mathfrak{L}_{q}$-norms of these orthogonal functions.
} 
\maketitle
\section{Introduction}
\label{intro}

The spatial localization/delocalization of quantum systems with high precision and its quantification play a central role in quantum mechanics. Presently the modern quantum technologies are making possible the actual realization of such experiments. The investigations have been driven by the possibility of practical applications including laser cooling and trapping, nanolithography and many other areas of atomic and molecular physics as well as classical and quantum information. While this uncertainty quantification was originally formulated in terms of variances and their generalization (the radial expectation values), they have later been successfully expressed with entropies of Fisher and Shannon types, and their generalization (particularly the R\'enyi entropies) in a much more appropriate manner.\\ 

The determination of these dispersion and entropy-like quantities for arbitrary quantum states is a formidable numerical task and certainly an impossible analytical task, except for a few solvable quantum-mechanical potentials which allow to model the mean field of numerous many-body systems. The spacial delocalization of physical systems with solvable quantum-mechanical potentials is an exciting, wide, rapidly progressing, cross-disciplinary topic, and that very nature makes it both attractive and hard to enter.\\ 

In this paper, we provide in part the tremendous advances made for quantifying the spacial delocalization of the multidimensional harmonic oscillator-like systems over the last two decades. 
The multidimensional harmonic oscillator is the most familiar confinement model to describe the structure of tightly bound systems with strong localization (see e.g. \cite{Zettili2009}). Undeniably, this harmonic system (i.e., a particle moving under the action of a quadratic potential) in one- and many dimensions is, together with the hydrogenic system, the most fundamental and far-reaching example in theoretical physics \cite{Landau1959,Han1994,Moshinsky1996,Bloch1997,Dong2011}. In spite of its simplicity, it plays a significant role from the early days of quantum mechanics \cite{Heisenberg1926} where it describes the behavior of systems close to an equilibrium position. For example, the low-energy spectrum of almost any quantum system is well described by an harmonic oscillator as can be shown, not only qualitatively but also quantitatively, by use of the elementary and supersymmetric quantum mechanics \cite{Goeppert1955,Gangopadhyaya2017}. Moreover, the energy spectrum of light and heavy atoms and molecules up until the proteines can be quite well fitted by means of the two parameters which characterize an effective multidimensional harmonic oscillator, the frequency and the dimensionality $D$. The multidimensional character of the system is justified not only for string theorists and cosmologists arguing that the best way to explain all forces of physics is via the idea of higher dimensions \cite{Wesson2006,Gallego2020}, but also because numerous physico-chemical phenomena of natural systems in our three-dimensional world can be explained via $D$-dimensional quantum mechanical quantities/objects with $D$ other than three in a much simpler and elegant manner \cite{Herschbach1993,Herschbach1996,Herschbach2020,Hooft2014}. \\

Nowadays there exists an increasing interest for multidimensional harmonic oscillators from neural networks \cite{Agliari2015}, fractional and quantum statistics \cite{Lepri2003,Asadian2013,Rovenchak2014,Armstrong2015,Kim2020}, classical and quantum information \cite{Yanez1994,Assche1995,Dehesa2001,Frieden2004,Zeilinger2017,Puertas2018,Krantz2019,Cabello2020,Plenio2020}, high-energy physics \cite{Hooft2014,Bures2015} to quantum chemistry and atomic and molecular physics  \cite{Dong2011,Benavides2014,Gadre1991,Aquino2010,Chang2002,Rantala2018,Pawlak2020}.\\ 

In this work we discuss the spatial spreading of the probability density $\rho(\mathbf{r}), \mathbf{r} \in \mathbb{R}_D$, of these harmonic systems by means of the dispersion-like measures (radial expectation values) and the information-theoretic entropies of local (Fisher information \cite{Frieden2004}) and global (Shannon and R\'enyi entropies \cite{Shannon1948,Shannon1993,Renyi1961}) character. Both dispersion and entropy measures reflect \textit{concentration} but their respective metrics for concentration are different. Unlike the former ones measure concentration only around a specific point (e.g., mean, origin,...) of $\mathbb{R}_D$, the latter ones quantify diffuseness of the density irrespective of the location(s) of concentration. The Fisher information, which is a gradient functional of the density, quantifies the concentration of the probability density around its nodes. The Shannon and R\'enyi entropies, which are logarithmic and power-like functionals of the density respectively, quantify the macroscopic facets of the density all over the multidimensional space. These entropy-like measures, which are the basic variables of the information theory of classical and quantum systems  \cite{Cover1991,Nielsen2000,Adesso2018}, are much more appropriate uncertainty measures than the (dispersion-like) Heisenberg measures. This is basically because the latter ones depend on a specific point of the domain of the density and give a large weight to the tails of the distribution (see e.g., \cite{Hilgevoord2002}), what is only true for some particular distributions such as those which fall off exponentially.\\

We show that these spreading quantities are analytically expressed in terms of the potential parameters and the hyperquantum numbers of the quantum oscillator states by use of a methodology based on the algebraic properties and techniques  of the special functions of mathematical physics \cite{Dehesa2001,Nikiforov1988}, such as e.g. the logarithmic potential and linearization methods of the orthogonal polynomials which control the corresponding wavefunctions. This methodology is, however, computationally demanding for the extreme states with high and very high values of the radial quantum number and/or the space dimensionality, since then the highly oscillatory nature of the integral functional kernels of the corresponding spreading measures renders ineffective the general analytical expressions and the conventional Gaussian quadrature algorithms. For these special states, the highly-excited Rydberg states and the high-dimensional (pseudo-classical) ones, we show that the weak* and strong degree-asymptotics \cite{Aptekarev1995,Buyarov1999,Aptekarev2010jcam} and the parameter-asymptotics \cite{Temme2017} of Laguerre and Gegenbauer polynomials allow us to find closed compact expressions for both the dispersion and entropy-like measures.  \\

The theoretical evaluation of the Shannon and R\'enyi entropies for the oscillator-like states is a formidable task \textit{per se} and respective to the determination of other uncertainty measures like the Heisenberg and Fisher information ones, not only analytically but also numerically. The latter is basically because a naive numerical evaluation using quadratures is not convenient due to the increasing number of integrable singularities when the radial hyperquantum number is increasing, which spoils any attempt to achieve reasonable accuracy even for rather small hyperquantum number \cite{Buyarov2004}. We should here mention for completeness that these entropies have been determined for the multidimensional hydrogenic states (see e.g. \cite{Dehesa2010,Puertas2018jstat,Toranzo2019ijqc} and for a number of quantum states of various interesting three-dimensional potentials of Hulthen, Yukawa and P\"{o}schl-Teller types \cite{Dehesa2006,Sun2013,Ikot2020}, among others.  \\

The purpose of this paper is three-fold: (a) to share with others this special-functions-based approach, which we have found to be successful for multidimensional harmonic oscillators; this material could be used by other researchers who want to develop delocalization outreach activities, (b) to provide a transparent step-by-step guide to other researchers who are interested in getting hands-on experience and some familiarity with uncertainty quantification in physical systems with solvable quantum potentials, and (c) to inspire others in the quantum uncertainty community to contribute by coming up with new strategies based on this approach. \\

The structure of the paper is the following. In Section \ref{sec:1} we briefly describe the quantum-mechanical wavefunctions and the associated probability densities of the $D$-dimensional harmonic oscillator (D-HO) in hyperspherical and Cartesian coordinates. In Section \ref{sec:2} we discuss the dispersion measures (the radial expectation values) of the D-HO and their associated Heisenberg uncertainty relations, which are then applied to Rydberg states and to high-dimensional oscillators. In Section \ref{sec:3} we analyze the entropy-like measures which quantify the multidimensional spreading of the D-HO in position and momentum spaces (the Fisher information, the Shannon and R\'enyi  entropies and the disequilibrium) and their associated uncertainty relations in both Cartesian and hyperspherical units. This is done by using techniques of potential theory and linearization of orthogonal polynomials. In Section \ref{sec:4} and \ref{sec:5} we explicitly study the dispersion and entropy-like measures of the Rydberg and high-dimensional states of the D-HO by means of the weak* and strong degree-asymptotics and the parameter-asymptotics of Laguerre and Gegenbauer polynomials, respectively. Finally, some conclusions and various open problems are given.

\section{The multidimensional harmonic oscillator: Basics}
\label{sec:1}

In this section we briefly describe in both Cartesian and hyperspherical coordinate systems the wavefunctions and the associated probability densities for the discrete stationary states of the $D$-dimensional harmonic system in both position and momentum spaces. This system corresponds to an isotropic harmonic oscillator described by the quantum-mechanical potential $V_{D}(r) =  \frac{1}{2}\omega^{2}r^{2}$, where $r=|\mathbf{r}|$ and the frequency $\omega$ is the oscillator strength. Later on, we give the associated probability densities for the stationary quantum states of the system. Atomic units (i.e., $\hbar = m_e = e = 1$) are used throughout the paper.
\subsection{The wavefunctions}
\label{subsec:1}

The time-independent non-relativistic equation of the $D$-dimensional ($D \geqslant 2$) harmonic system is given by the Schr\"{o}dinger equation
\begin{equation}
\label{eqI_cap1:ec_schrodinger}
\left( -\frac{1}{2} \vec{\nabla}^{2}_{D} + V_{D}(r)\right) \Psi \left(\mathbf{r} \right) = E \Psi \left(\mathbf{r} \right),
\end{equation}
where $\vec{\nabla}_{D}$ denotes the $D$-dimensional gradient operator, and the  position vector $\mathbf{r} =(x_1 ,  \ldots  , x_D) = (r,\theta_1,\theta_2,\ldots,\theta_{D-1})$ in Cartesian and hyperspherical units, respectively. Moreover, $r \equiv |\mathbf{r}| = \sqrt{\sum_{i=1}^D x_i^2}
\in [0 , \: +\infty)$  and $x_i =  r \left(\prod_{k=1}^{i-1}  \sin \theta_k
\right) \cos \theta_i$ for $1 \le i \le D$
and with $\theta_i \in [0 , \pi), i < D-1$, $\theta_{D-1} \equiv \phi \in [0 ,  2
\pi)$. 

The wavefunctions in Cartesian coordinates are known (see e.g. \cite{Puertas2018}) to be characterized by the Cartesian quantum numbers ${\{n_i\}} \equiv (n_1, n_2, ...,n_D)$ and described by the energetic eigenvalues
\begin{equation}
\label{HOEL}
E_{\{n_i\}} = \left(N + \frac{D}{2}\right) \omega, \quad\text{with}\quad N = \sum_{i=1}^{D}n_{i}\,; \quad n_i=0,1,2,\ldots
\end{equation}
and the associated eigenfunctions 
\begin{equation}
\label{HOEF}
\Psi_{\{n_i\}}(\mathbf{r}) = \mathcal{N}\, e^{-\frac{1}{2}\alpha'(x_{1}^{2}+\ldots+x_{D}^{2})}H_{n_{1}}(\sqrt{\alpha'}\, x_{1})\cdots H_{n_{D}}(\sqrt{\alpha'}\, x_{D}), 
\end{equation}	
where $\alpha' = \omega^{\frac{1}{4}}$, $H_{n_i}(x)$ denotes the Hermite polynomial of degree $n_i$ orthogonal with respect the weight function $\omega(x) = e^{-x^{2}}$ in $(-\infty, \infty)$, and $\mathcal{N}$ denotes the normalization constant
\[
\mathcal{N} = \frac{1}{\sqrt{2^{N}n_{1}!n_{2}!\cdots n_{D}! }}\left(\frac{\alpha'}{\pi}\right)^{D/4}.
\]
On the other hand, working in hyperspherical coordinates, it is known that the wavefunctions are described \cite{Yanez1994} by the energetic eigenvalues 
\begin{equation}
\label{eq:oscillenerg}
E_{n,l}= \left(\eta+ \frac{3}{2}\right)\,\omega   = \left(2n_r+l+\frac{D}{2}\right) \omega, \quad \text{with}\quad \eta = n+ \frac{D-3}{2},\quad n = 2n_r+l 
\end{equation}
and the associated eigenfunctions 
\begin{eqnarray}
\label{eq:wavpos}
\label{eq:wavmom}
\Psi_{n_r,l,\{\mu\}}(\mathbf{r}) &=& \left[\frac{2n_r!\,\omega^{l+\frac{D}{2}}}{\Gamma(n_r+l+\frac{D}{2})} \right]^{\frac{1}{2}}r^{l}e^{-\frac{\omega\,r^{2}}{2}}\mathcal{L}^{(l+D/2-1)}_{n_r}(\omega\, r^{2})\,\mathcal{Y}_{l,\{\mu\}}(\Omega_{D-1}),\\
\hat{\Psi}_{n_r,l,\{\mu\}}(\mathbf{p}) &=& \left[\frac{2n_r!\,\omega^{-l-\frac{D}{2}}}{\Gamma(n_r+l+\frac{D}{2})} \right]^{\frac{1}{2}}p^{l}e^{-\frac{p^{2}}{2\omega}}\mathcal{L}^{(l+D/2-1)}_{n_r}\left(\frac{p^{2}}{\omega}\right)\, \mathcal{Y}_{l,\{\mu\}}(\Omega_{D-1})
\end{eqnarray}
in position and momentum spaces, respectively. Note that the momentum wavefunctions are the Fourier transform of the position ones. The symbol $\mathcal{L}_{n}^{(\alpha)}(x)$ denotes the orthogonal Laguerre polynomials \cite{Olver2010} with respect to the weight function $\omega_\alpha(x)=x^{\alpha} e^{-x}, \, \alpha= l+\frac D2-1,$ on the interval $\left[0,\infty \right)$.
The angular part of the eigenfunctions are the hyperspherical harmonics, $\mathcal{Y}_{l,\{\mu\}}(\Omega_{D-1})$, defined \cite{Yanez1994,Avery2006,Coletti2013} by
\vspace{-1.1mm}
\begin{equation}\label{armonico}
{\cal{Y}}_{l,\left\lbrace \mu \right\rbrace}(\Omega_{D-1})=\frac{1}{\sqrt{2 \pi}} e^{im\theta_{D-1}} \prod^{D-2}_{j=1} \tilde{C}_{\mu_{j}-\mu_{j+1}}^{(\alpha_j+\mu_{j+1})}(\cos \theta_j) \left( \sin \theta_j\right)^{\mu_{j+1}},
\end{equation}
with $2\alpha_j=D-j-1$ and $\tilde{C}^{(\lambda)}_n(x)$, $\lambda>-\frac{1}{2}$, denotes the Gegenbauer polynomial of degree n and parameter $\lambda$ which satisfies the orthonormalization condition
\vspace{-1mm}
\begin{equation}\label{ortoCortonormales}
\int_{-1}^1\tilde{C}^{(\lambda)}_n(x)\tilde{C}^{(\lambda)}_m(x) \omega_\lambda(x)dx=\delta_{mn},
\end{equation}
where the weight function is given by $\omega_\lambda(x)=\left( 1-x^2 \right)^{\lambda-\frac{1}{2}}.$\\

Note that the wavefunctions are duly normalized so that  $\int \left| \Psi_{\eta,l, \left\lbrace \mu \right\rbrace }(\mathbf{r}) \right|^2 d\mathbf{r} = \int \left| \Psi_{\eta,l, \left\lbrace \mu \right\rbrace }(\mathbf{p}) \right|^2 d\mathbf{p} = 1$, where the $D$-dimensional volume elements are $d\mathbf{r} = r^{D-1} dr\,d\Omega_{D-1}$ and  $d\mathbf{p} = p^{D-1} dp\,d\Omega_{D-1}$ respectively, with the generalized solid angle element
\[
d\Omega_{D-1}=\left(\prod_{j=1}^{D-2} (\sin \theta_j)^{2 \alpha_j} d\theta_j\right) d\theta_{D-1},
\]
and we have taken into account the normalization to unity of the hyperspherical harmonics given by $\int |\mathcal{Y}_{l,\{\mu \}}(\Omega_{D-1})|^{2}d\Omega_{D-1} = 1$.

Let us also highlight that the oscillator wavefunctions in the hyperspherical coordinate system $(r,\theta_1,\theta_2,\ldots,\theta_{D-1})$ are characterized by the $D$ hyperquantum integer numbers $(n_r,l,\left\lbrace \mu \right\rbrace)\equiv (n_r,\mu_1,\mu_2,\ldots,\mu_{D-1})$ with the values $n_r=0,1,2,\ldots,\, l=0,1,2,\ldots,$ and $l\,\equiv\, \mu_{1} \geq \mu_{2} \geq \ldots \geq |\mu_{D-1}| \equiv |m|$.

\subsection{The probability densities}
\label{subsec:2}

The probability density of the $D$-dimensional isotropic harmonic oscillator in both Cartesian and hyperspherical coordinate systems is given by the modulus squared of the corresponding eigenfunctions, obtaining the expressions
\begin{equation}
\label{HOPD}
\rho_{\{n_i\}}(\mathbf{r}) = |\psi_{\{n_i\}}(\mathbf{r})|^{2} = \mathcal{N}^{2}\, e^{-\alpha'(x_{1}^{2}+x_{2}^{2}+\ldots+x_{D}^{2})}H_{n_{1}}^{2}(\sqrt{\alpha'}\, x_{1})\cdots H_{n_{D}}^{2}(\sqrt{\alpha'}\, x_{D}),
\end{equation}
and 
\begin{eqnarray}
\label{eq:densposhyper}
\rho_{n_r,l,\{\mu \}}(\mathbf{r}) &=& \frac{2n_r!\,\omega^{l+\frac{D}{2}}}{\Gamma(n_r+l+\frac{D}{2})}r^{2l}e^{-\omega\, r^{2}}\left[\mathcal{L}^{(l+D/2-1)}_{n_r}(\omega\, r^{2})\right]^{2}\times |\mathcal{Y}_{l,\{\mu\}}(\Omega_{D-1})|^{2}\nonumber\\
&=& 2\,\omega^{\frac{D}{2}}\tilde{r}^{1-\frac{D}{2}}\omega_{l+D/2-1}(\tilde{r})\,[\tilde{{\cal{L}}}_{n}^{(l+D/2-1)}(\tilde{r})]^{2} \times |\mathcal{Y}_{l,\{\mu\}}(\Omega_{D-1})|^{2} \\
&\equiv & \rho_{n_r,l}(r) \times \rho_{l,\{\mu \}}(\Omega_{D-1})  
\end{eqnarray}
(where $\tilde{r}= \omega \,r^{2}$ and $\omega_{\alpha}(x) =x^{\alpha}e^{-x}$ is the weight function with respect to which the Laguerre polynomials $\left\{\mathcal{L}_{n}^{(\alpha)}(x), \tilde{{\cal{L}}}_{n}^{(\alpha)}(x)\right\}$, are orthogonal and orthonormal, respectively) for the position probability density, and the expressions 
\begin{align}
\label{HOMPD}
\gamma_{\{n_i\}}(\mathbf{p}) & = \mathcal{\tilde{N}}^{2} e^{-\frac{1}{\alpha'}(p_{1}^{2}+p_{2}^{2}+\ldots+p_{D}^{2})}H_{n_{1}}^{2}\left(\frac{ p_{1}}{\sqrt{\alpha'}}\right)\cdots H_{n_{D}}^{2}\left(\frac{ p_{D}}{\sqrt{\alpha'}}\right)
= \alpha'^{-D}\rho_{N}\left(\frac{\vec{p}}{\alpha'}\right)
\end{align}
with the normalization constant 
\[
\mathcal{\tilde{N}}  = \frac{1}{\sqrt{2^{N}n_{1}!n_{2}!\cdots n_{D}! }}\left(\frac{1}{\pi\alpha'}\right)^{D/4}.
\]
and
\begin{eqnarray}
\label{eq:densmom}
\gamma_{n_r,l,\{\mu \}}(\mathbf{p}) &=& \frac{2 n_r!\,\omega^{-l-\frac{D}{2}}}{\Gamma(n_r+l+\frac{D}{2})}p^{2l}e^{-\frac{p^{2}}{\omega}}\left[\mathcal{L}^{(l+D/2-1)}_{n_r}\left(\frac{p^{2}}{\omega}\right)\right]^{2} |\mathcal{Y}_{l,\{\mu\}}(\Omega_{D-1})|^{2} \nonumber \\
&=& \frac{1}{\omega^{D}}\,\rho\left(\frac{\vec{p}}{\omega}\right),
\end{eqnarray}
for the momentum probability density, respectively.

\section{Dispersion measures, Heisenberg uncertainty, Rydberg states and high dimensional oscillators}
\label{sec:2}

\subsection{The radial expectation values}
\label{subsec:3}

In this section we obtain the radial expectation values of the $D$-dimensional harmonic state $(n_r,l,\{\mu\})$ in both position and momentum spaces, denoted by $\langle r^{k}\rangle$ and $\langle p^{k}\rangle$, respectively. Then, we apply the general resulting expressions to some relevant cases such as the Rydberg oscillator states and the high-dimensional oscillator systems. Finally, the associated Heisenberg uncertainty products are explicitly shown.\\
We start with the expressions given by Equations (\ref{eq:densposhyper}) and (\ref{eq:densmom}) of the position and momentum probability densities of the system, respectively, obtaining the expressions \cite{Puertas2017,Zozor2011} 
\begin{eqnarray}
\label{1}
\langle r^{k}\rangle &=& \int_{\mathbb{R}_D} r^{k}\rho_{n_r,l,\{\mu \}}(\mathbf{r}) \,d\mathbf{r} 
= \frac{\Gamma(n_r+1)\,\omega^{-k/2}}{\Gamma(n_r+l+D/2)}\int_{0}^{\infty} x^{\alpha+\frac{k}{2}}e^{-x}[\mathcal{L}^{(\alpha)}_{n_r}(x)]^{2}\, dx\nonumber\\
&=& \omega^{-\frac{k}{2}}\frac{\Gamma\left(l+\frac{D+k}{2}\right)}{\Gamma\left(l+\frac{D}{2}\right)} \, {}_3F_{2} \left(-n_r,-\frac{k}{2},\frac{k}{2}+1;l+\frac{D}{2},1;1\right)\\
&=& \omega^{-\frac{k}{2}}\frac{n_r!}{\Gamma\left(n_r+l+\frac{D}{2}\right)} \, \sum_{i=0}^{n_r} \binom{k/2}{n_r-i}^2\,\frac{\Gamma\left(l+\frac{D+k}{2}+i\right)}{i!}
\label{eq:defhf0}
\end{eqnarray}
(with $x=\omega\, r^{2},\,  \alpha=l+D/2-1,\,k >-D-2l$) for the radial expectation values in position space, and 
\begin{eqnarray}
\langle p^{k}\rangle &=&\int_{\mathbb{R}_D} p^{k}\gamma_{n_r,l,\{\mu \}}(\mathbf{p})\, d\mathbf{p} 
= \omega^{k} \,\langle r^{k}\rangle 
\label{eq:defhf}
\end{eqnarray}
for the radial expectation values in momentum space. Note that we have made profit of the unity normalization of the hyperspherical harmonics mentioned above, and we have used the symbol $_3F_2(1)$ which denotes the value of the generalized hypergeometric function ${}_{3}F_{2}(z)$ at $z=1$. This function is the generalized hypergemetric series \cite{Olver2010} given by
\begin{equation}
_{3}F_2\left(a_1,a_2,a_3;b_1,b_2;z\right) = \sum_{j=0}^{\infty} \frac{(a_{1})_{j}(a_{2})_{j}(a_{3})_{j}}{(b_{1})_{j}(b_{2})_{j}}\frac{z^{j}}{j!}, 
\end{equation}
(where $(a)_j = \Gamma(a+j)/\Gamma(a)$ denotes the well-known Pochhammer symbol) which is a terminating series when the first one or more of the top parameters is a nonnegative integer, so that it reduces to a polynomial in $z$. Thus, ${}_3F_{2}(1)$ is a terminating hypergeometric function (so, a polynomial) evaluated at $z=1$; in fact, it is a discrete polynomial of dual Hahn type \cite{Suslov2009,Nikiforov1991}. From Equation (\ref{eq:defhf0}) various relevant properties for the radial expectation values in position space follow, such as the Kramer-like three-term recurrence relation (see also \cite{Dong2011,Ray1988})
\begin{equation}\label{eq:TTRR}
(k+2)\,\omega^2\,\langle r^{k+2}\rangle= (k+1)\,\omega\,(2n+D)\langle r^{k}\rangle+k\left(\frac{k^2-D^2}{4}-(l-1)(l+k-1)\right)\langle r^{k-2}\rangle,
\end{equation} 
where $n=2n_r+l$ as mentioned above. Note that this expression yields recurrence relations either between even moments or between odd moments only. For instance, with the initial conditions
\begin{equation}\label{eq:evm2}
\langle r^{0}\rangle = 1\quad  \text{and} \quad \langle r^{-2}\rangle = \frac{1}{L+1/2} \omega 
\end{equation}
(which easily follow from Equation (\ref{eq:defhf0})), the recurrence relation (\ref{eq:TTRR}) gives the explicit expression for all the even moments, such as e.g.
\begin{eqnarray}
\label{eq:expecrminus4}
\langle r^{-4}\rangle &=& \frac{\eta+\frac{3}{2}}{\left(L - \frac{1}{2}\right) \left(L + \frac{1}{2}\right) \left(L + \frac{3}{2}\right)} \, \omega^2\\
\label{eq:expecr2}
\langle r^{2}\rangle &=&  \left(\eta + \frac{3}{2}\right)\omega^{-1} = \left(2n_r+l+\frac{D}{2}\right)\omega^{-1}\\  
\label{eq:expecr4}
\langle r^{4}\rangle &=& \frac{1}{2}\left[3\left(\eta + \frac{3}{2}\right)^2 - \left(L - \frac{1}{2}\right)\left(L + \frac{3}{2}\right)\right]  \omega^{-2},
\end{eqnarray}
where we should keep in mind that $L=l+\frac{D-3}{2}$, $\eta = n + \frac{D-3}{2} = 2\,n_{r}+l+\frac{D-3}{2} = 2\,n_{r}+L$, and  $n_{r}=0,1,2,\ldots$
On the other hand, with the initial conditions $(\langle r\rangle, \langle r^{-1}\rangle)$, 
the relation in Equation (\ref{eq:TTRR}) gives all the odd moments. See also \cite{Dong2011,Ray1988,Louck1985,Romera2006} for further values and properties. Moreover, it is also possible to find  the reflection formula \cite{Dong2011,Suslov2009}
\begin{equation}
\langle r^{-k+2}\rangle = \frac{\Gamma\left(l+\frac{D-k}{2}-1\right)}{\Gamma\left(l+\frac{D+k}{2}\right)} \,   \langle r^{k}\rangle,
\end{equation}
and
\begin{equation}
\langle r^{-3}\rangle = \frac{4\,\omega^2}{\left(D-1+2\,l\right)\left(D-3+2\,l\right)} \,   \langle r\rangle
\end{equation}
The corresponding reflection formula for momentum expectation values easily follows from Eq. (\ref{eq:defhf}).\\

However, the computation of the expectation values is a formidable task for the oscillator states of Rydberg type (high $n_r$, fixed $D$) and for the high-dimensional oscillator systems. In these cases the highly oscillatory nature of the corresponding integrands (see Equation (\ref{eq:defhf0})) creo que deberíamos nombrar la primera expresión de $\langle r^{k}\rangle$, justo la que se encuentra por encima de la (14) porque es donde aparece la integral renders Gaussian quadrature ineffective as the number of quadrature points grows linearly with $n$ and evaluation of the involved high-degree polynomials are subject to round-off errors. The calculation of the expectation values in these extreme cases require asymptotical methodologies of highbrow character which supply a closed compact expression, as shown in the following. 

\subsection{Rydberg states}
\label{subsec:4}

To calculate the position and momentum expectation values of the Rydberg states (i.e., states with a high or very high radial quantum number $n_r$) for the $D$-dimensional harmonic oscillator, we use a specific methodology based on the weak*- asymptotics of Laguerre polynomials \cite{Buyarov1999,Aptekarev2010jpa} which has allowed us to obtain the expressions  
\begin{equation}
\label{eq:asympr1}
\langle r^{k}\rangle \simeq  (a\,n_r)^{\frac{k}{2}}\,
\ _2F_{1}\left(-\frac{k}{2},\frac{1}{2},1;\frac{2}{s^{2}}(-1+s^{2}-\sqrt{1-s^{2}})  \right) \,\omega^{-\frac{k}{2}}
\end{equation}
where $a=\frac{2}{1-s}(1+\sqrt{1-s^{2}})$ and $\lim\limits_{n_r\rightarrow \infty}\frac{l}{n_r} =s \in [0,1)$. Then,
when $l$ is uniformly bounded (so that $s=0$ and $a=4$) we get
\begin{eqnarray}
\label{eq:asympr2}
\langle r^{k}\rangle &\simeq& (4\,n_r)^{\frac{k}{2}}\frac{\Gamma\left(\frac{1+k}{2}\right)}{\sqrt{\pi}\,\Gamma\left(1+\frac{k}{2}\right)}\, \omega^{-\frac{k}{2}} 
\end{eqnarray} 
with $k >-1$, which gives the following first position expectation values 
\begin{equation}
\label{eq:somers}
\langle r^{0}\rangle =1\, , \quad \langle r\rangle \simeq \frac{2}{\pi}\sqrt{4\,n_r} \,\omega^{-\frac{1}{2}} , \quad \langle r^{2}\rangle \simeq \,2\,n_r\, \omega^{-1}, \quad \quad \langle r^{4}\rangle \simeq  6\,n_r^2\, \omega^{-2}.
\end{equation}
Then, the corresponding momentum expectation values $\langle p^{k}\rangle =  \langle r^{k}\rangle\,\omega^{k} $ are
\begin{equation}
\label{eq:someps}
\langle p^{0}\rangle =1\, , \quad \langle p\rangle \simeq \frac{2}{\pi}\sqrt{4\,n_r} \,\omega^{\frac{1}{2}} , \quad \langle p^{2}\rangle \simeq  2\,n_r\, \omega, \quad \langle p^{4}\rangle \simeq  6\,n_r^2\,\omega^{2}.
\end{equation}
Note that Equation (\ref{eq:asympr2}) gives the dominant term for the Rydberg expectation values $\langle r^{k}\rangle$ with $k >-1$ which does not depend on the oscillator dimensionality. To extend these results for $k\leq -1$, to explicitly have the dimensionality dependence and to uncover terms beyond the dominant one, it is necessary to use the strong asymptotics of the Laguerre polynomials \cite{Aptekarev1995,Aptekarev2010jcam,Aptekarev1994,Dehesa1998} what is an open problem. Let us also point out that e.g., Equations (\ref{eq:evm2})-(\ref{eq:expecr4}) and Equation (\ref{eq:somers}) are mutually consistent.

\subsection{High-dimensional oscillators}
\label{subsec:5}

To compute the expectation values for harmonic oscillators with high and very-high dimensionality $D$ it is most convenient to use a modern methodology based on the parameter-asymptotics \cite{Temme2017} (see also Corollary 1 of Appendix A in \cite{Puertas2017}) of the integral functionals of Laguerre polynomials $\mathcal{L}^{(\alpha)}_{n}(x)$ involved in Equation (\ref{eq:defhf0}) lo mismo que en la página anterior, creo que debería nombrarse la primera expresión de $\langle r^{k}\rangle$, obtaining the position expectation values
\begin{align}
\label{7}
\langle r^{k}\rangle 
&\simeq \sqrt{2\pi}\,\,e^{-\alpha}\frac{\alpha^{\alpha+n_r+(k+1)/2}}{\Gamma(n_r+l+D/2)}\,\omega^{-k/2}, \qquad \alpha \to \infty
\end{align}
(with $\alpha=l+D/2-1$) for the harmonic states with fixed $l$. The corresponding momentum expectation values follow from Equation (\ref{eq:defhf}). Then, with the first order ($z \to \infty$)-asymptotical expansion of the Gamma function \cite{Olver2010}, $\Gamma(z)\sim \sqrt{2\pi}z^{z-1/2}e^{-z}$, one easily has that 
\begin{equation}
\label{9}
\langle r^{k}\rangle 
\simeq \left(\frac{D}{2\,\omega}\right)^{\frac{k}2};\quad \langle p^{k}\rangle 
\simeq \left(\frac{D\,\omega}{2}\right)^{\frac{k}2}, \qquad \text{for}\qquad D>> 1.
\end{equation}
Note that, since the dependence on the quantum numbers is lost, the intrinsic quantum-mechanical structure of the system gets hidden in the limit $D \to \infty$,  what is a manifestation of the closeness to the (pseudo-) classical situation. In addition, we observe the existence of a characteristic length for this system, $r_c = \left(\frac{D}{2\omega}\right)^{\frac{1}2}$ at this pseudoclassical limit since then we have that $\langle r\rangle \to r_c$ and $\langle r^{k}\rangle \to r_c^k$. Moreover, from Equation (\ref{eq:oscillenerg}) the energy behaves  as $E \to \omega \,\frac{D}{2} = \omega^2 r_c^2$.  This characteristic length corresponds to the distance at which the effective potential becomes a minimum and the ground state probability distribution has a maximum \cite{Ray1988}. Therefore, the $D$-dimensional oscillator in the $D \to \infty$ can be viewed as a particle moving in a classical orbit of radius $r_c$ with energy $E = \omega^2 r_c^2$ and angular momentum $L = D/2$.

\subsection{Heisenberg uncertainty products}
\label{subsec:6}

From Equations (\ref{eq:defhf0}) and (\ref{eq:defhf}) one obtains the generalized Heisenberg uncertainty product \begin{equation}\label{eq:genheispro}
\langle r^{k}\rangle\,\langle p^{k}\rangle	= \left[\frac{\Gamma\left(l+\frac{D+k}{2}\right)}{\Gamma\left(l+\frac{D}{2}\right)} \, {}_3F_{2} \left(-n_r,-\frac{k}{2},\frac{k}{2}+1;l+\frac{D}{2},1;1\right) \right]^2,
\end{equation}
which does not dependence on the oscillator strength $\omega$ as expected because of the homogenous property of the oscillator potential \cite{Sen2006}. Moreover, for $k=2$ it gives the familiar relation  
\begin{equation}
\label{eq:heispro}
\langle r^{2}\rangle\langle p^{2}\rangle = \left(2n_{r}+l+\frac{D}{2}\right)^{2} = \frac{D^{2}}{4}\left\{1+\frac{1}{D}(8n_r+4l)+\frac{1}{D^{2}}[4 (2n_r+l)^2] \right\},
\end{equation}
which fulfills not only the celebrated Heisenberg uncertainty relation $\langle r^{2}\rangle\langle p^{2}\rangle \geq  \frac{D^{2}}{4}$ of the general $D$-dimensional quantum systems but, since $n_r \geq 0$, also the tighter Heisenberg uncertainty relation of the quantum systems subject to any spherically-symmetric $D$-dimensional potential given \cite{Sanchez2006} by
\begin{equation}
\label{eq:centrpotheis1}
\langle r^{2}\rangle\langle p^{2}\rangle \geq  \left(l+\frac{D}{2} \right)^{2}.
\end{equation}
From Equation (\ref{eq:heispro}) one observes that this inequality gets saturated by the (nodeless) ground state of the $D$-dimensional harmonic system. In addition note that, according to Equation (\ref{eq:asympr2}), the generalized Heisenberg uncertainty product in Equation (\ref{eq:genheispro}) for Rydberg states of an oscillator system with a given dimensionality gets simplified in the framework of the the weak*-asymptotics of Laguerre polynomials as 
\begin{equation}
\label{eq:asymproduct1}
\langle r^{k}\rangle\langle p^{k}\rangle \simeq (4\, n_r)^{k}\, \pi^{-1}\left[\frac{\Gamma\left(\frac{1+k}{2}\right)}{\Gamma\left(1+\frac{k}{2}\right)}\right]^2
\end{equation}
when $n_r\rightarrow \infty$ and $l$ uniformly bounded. In the particular case $k=2$ this product reduces to
\begin{equation}
\label{eq:asymproduct2}
\langle r^{2}\rangle\langle p^{2}\rangle \simeq 4\, n_r^{2}.
\end{equation}
The explicit dependence on the dimensionality can be obtained by means of the strong asymptotics of the Laguerre polynomials \cite{Aptekarev1994,Aptekarev1995,Aptekarev2010jcam}, as already mentioned.

Finally, from Equation (\ref{9}), one has that the generalized Heisenberg uncertainty product in Equation (\ref{eq:genheispro}) has the simpler form
\begin{equation}
\label{eq:heilike}
\langle r^k\rangle\langle p^k\rangle =\left(\frac D2\right)^k \left(1+ \mathcal{O}\left(\frac{1}{D^{}} \right)\right),
\end{equation}
for the generalized Heisenberg uncertainty product of the high-dimensional oscillator-like systems. Terms in this expression beyond the first order can be obtained by using the ($\alpha \to \infty$)-asymptotics \cite{Temme2017} (see also Corollary 1 of Appendix A in \cite{Puertas2017}) of the abovementioned integral functionals of Laguerre polynomials.

\section{The entropy-like measures and entropic uncertainty}
\label{sec:3}

In this section we obtain the position and momentum entropy-like of Fisher, Shannon and R\'enyi types and the disequilibrium for any $D$-dimensional harmonic state $(n_r,l,\{\mu\})$ denoted by $F[\rho], S[\rho]$, $R_q[\rho]$ and $\mathcal{D}[\rho]$, respectively. Then, the associated position-momentum entropic uncertainty products are explicitly given. The convenience to use the hyperspherical units or the Cartesian ones is discussed in each case. \\

Shortly, let us advance that (a) the Fisher information can be expressed directly in terms of the  hyperspherical quantum numbers, mainly because of the close similarity of its gradient-functional form and the multidimensional kinetic energy of the oscillator-like system, and (b) the Shannon and R\'enyi entropies cannot be explicitly expressed by hyperspherical quantum numbers, basically because these logarithmic and power-like quantities naturally depend on some entropy-like integral functionals of the Laguerre and Gegenbauer orthogonal polynomials whose analytical evaluation are not yet known despite so many efforts (see e.g. \cite{Yanez1994,Dehesa2001,Bhattacharya1998,Ghosh2000}). Below, we illustrate this difficult task by analyzing in detail the evaluation of the simplest second-order R\'enyi entropy, the disequilibrium, in both hyperspherical and Cartesian coordinate systems. Moreover, we also show later on that these global entropies can be analytically determined in Cartesian hyperquantum units.

\subsection{The Fisher information}
\label{subsec:7}

The Fisher information for $D$-dimensional harmonic state $(n_r,l,\{\mu\})$ with the position probability density $\rho(\mathbf{r}) \equiv \rho_{n_r,l,\{\mu \}}(\mathbf{r})$ given by Equation (\ref{eq:densposhyper}), is defined by
\begin{eqnarray}
\label{eq:posFisher}
F[\rho_{n_r,l,\{\mu \}}] &:=& \int_{\mathbb{R}_D}\frac{|\mathbf{\nabla}_{D}\,\rho_{n_r,l,\{\mu \}}(\mathbf{r})|^{2}}{\rho_{n_r,l,\{\mu \}}(\mathbf{r})}\, d\mathbf{r}.
\end{eqnarray}
This quantity is a local spreading measure of the state's density because it is a gradient functional of it. It quantifies the concentration of the probability around the density nodes. Moreover, it describes a local uncertainty measure so that the higher this quantity is, the more localized is the density, the smaller is the uncertainty and the higher is the accuracy in predicting the localization of the particle. The corresponding
quantity for the momentum probability density $\gamma(\mathbf{p}) \equiv \gamma_{n_r,l,\{\mu \}}(\mathbf{p})$ will be denoted by $F \left[ \gamma \right]$.\\

To calculate these two Fisher informations we first take into account that the oscillator potential is a central potential and then we use the radial expectation values $\langle r^{k}\rangle$ and $\langle p^{k}\rangle$ with $k=−2$ and $2$ given by Equation (\ref{eq:evm2}), (\ref{eq:expecr2}) and (\ref{eq:defhf}), obtaining 
\begin{eqnarray}
\label{eq:fishinf1}
F[\rho_{n_r,l,\{\mu \}}] &=& 4\langle p^{2}\rangle -2|m|(2l+D-2)\langle r^{-2}\rangle = 4\left( \eta -|m|+\frac{3}{2} \right)\omega \\
&=& 4\left( 2\,n_{r}+l -|m|+\frac{D}{2} \right)\omega
\end{eqnarray}
in position space, and
\begin{eqnarray}
\label{eq:fishinf2}
F \left[ \gamma_{n_r,l,\{\mu \}} \right] &=& 4\langle r^{2}\rangle -2|m|(2l+D-2)\langle p^{-2}\rangle =4\left( \eta -|m|+\frac{3}{2} \right)\omega^{-1} \\
&=& 4\left( 2\,n_{r}+l -|m|+\frac{D}{2} \right)\omega^{-1}
\end{eqnarray}
in momentum space, which were previously found for three dimensional oscillators \cite{Romera2005}. Note that these expressions boil down in the ground state (i.e., $n_r = l = \{\mu\} = 0$) to the values
\begin{equation}
\label{eq:fishinf}
F[\rho_{0,0,\{0 \}}] = 4\langle p^{2}\rangle = 2D\,\omega; \quad F \left[ \gamma_{0,0,\{0 \}}\right]  = 4\langle r^{2}\rangle = 2D\,\omega^{-1}, 
\end{equation}
which saturate the celebrated Stam inequalities for general quantum systems; i.e. the inequalities $F\left[ \rho\right] \leq 4 \left\langle p^2 \right\rangle$ and $ F\left[ \gamma \right] \leq 4 \left\langle r^2 \right\rangle$.
Moreover,  for all stationary D-HO states $(n_r,l,\{\mu\})$ we have that the position-momentum Fisher-information-based uncertainty product is given by
\begin{equation}\label{eq:fisherpro}
F[\rho_{n_r,l,\{\mu \}}]\times F \left[ \gamma_{n_r,l,\{\mu \}} \right] = 16\left( 2\,n_{r}+l -|m|+\frac{D}{2} \right)^2,
\end{equation}
which satisfies and saturates not only the position-momentum Fisher-information-based uncertainty relation
\begin{equation}
F[\rho]\times F \left[ \gamma \right] \geq	4 D^2
\end{equation}
(valid for both the one-dimensional harmonic oscillators \cite{Dehesa2006} and the general $D$-dimensional quantum systems with real-valued position or momentum wavefunctions \cite{Sanchez2011}), but also the position-momentum Fisher-information-based uncertainty relation
\begin{equation}
\label{eq:uncerrelfisher2}
F[\rho]\times F \left[ \gamma \right] \geq 16\left(l+\frac{D}{2}\right)^{2} \left[ 1- \frac{2\,|m|}{2l+D-2}  \right]^{2},
\end{equation}
which holds for all the stationary states of $D$-dimensional single-particle systems with a central potential \cite{Sanchez2006,Dehesa2007jpa}.
From Equation (\ref{eq:fisherpro}) one observes that the last two inequalities get saturated by the (nodeless) ground state of the D-HO system.

\subsection{The Shannon entropies}
\label{subsec:8}

The Shannon entropy for the generic $D$-dimensional oscillator-like state with the position probability density $\rho(\mathbf{r})$ given by Equations (\ref{HOPD}) and (\ref{eq:densposhyper}) in Cartesian and hyperspherical units, respectively, is defined \cite{Shannon1948} by
\begin{eqnarray}
\label{eq:posShannon}
S[\rho] &:=& -\int_{\mathbb{R}_D} \rho(\mathbf{r})\log \rho(\mathbf{r})\, d\mathbf{r}.
\end{eqnarray}
This quantity is a global spreading measure of the state's density which does not depend on any specific point of its multidimensional domain. It quantifies the total spreading of the density. Moreover, it describes a global uncertainty measure so that the higher this quantity is, the more delocalized is the density, the higher is the uncertainty and the smaller is the accuracy in predicting the localization of the particle. The corresponding quantity for the probability density $\gamma(\mathbf{p})$ in momentum space will be denoted by $S[\gamma]$.\\

The computation of the Shannon entropy for an arbitrary oscillator-like state in hyperspherical units has not yet been explicitly obtained by means of the state's hyperquantum numbers. Basically, the reason is that (a) this physical entropy depends on the entropy-like functional of the Laguerre polynomials and the entropy-like functional of the hyperspherical harmonics which ultimately can be expressed in terms of the entropy-like functional of the Gegenbauer polynomials) \cite{Yanez1994,Dehesa2001,Yanez1999,Sanchez2000,Dehesa2007jmp}; and (b)  these two mathematical Laguerre and Gegenbauer entropies, which are given by
\begin{equation}\label{eq:polentropy}
\mathcal{E}(y_n) = -\int^{b}_{a} \omega_{\alpha}(x)\, y^{2}_{n}(x) \log y^{2}_{n}(x) dx,
\end{equation}
(where $y_{n}(x) = \mathcal{L}^{(\alpha)}_{m}(x),\, \mathcal{C}^{(\alpha)}_{m}(x)$ denote the Laguerre and Gegenbauer polynomials orthogonal with respect to the weight function $\omega_{\alpha}(x)=x^{\alpha} e^{-x}, \,(1-x^2)^{\alpha-1/2}$ defined on the interval $(a,b) = (0,\infty), (-1,1)$, respectively), have not been analytically found in terms of the polynomial parameters up until now despite so many efforts \cite{Yanez1994,Dehesa2001,Dehesa1997,Buyarov2000,Vicente2007}. Moreover, the numerical determination of these quantities is not at all trivial \cite{Buyarov2004} as mentioned above.\\

Indeed, \textbf{in hyperspherical coordinates} the position Shannon entropy for an arbitrary $D$-dimensional oscillator-like state characterized by the probability density $\rho_{n_r,l,\{\mu \}}(\mathbf{r})$ is given, according to Equation (\ref{eq:densposhyper}) and (\ref{eq:posShannon}), by
\begin{eqnarray}
\label{eq:posShannonos}
S[\rho_{n_r,l,\{\mu \}}] &=& -\int_{\mathbb{R}_D} \rho_{n_r,l,\{\mu \}}(\mathbf{r})\log \rho_{n_r,l,\{\mu \}}(\mathbf{r})\, d\mathbf{r}\\
\label{eq:shannontotal}
&=& S[\rho_{n_r,l}] + S[\rho_{l,\{\mu \}}],
\end{eqnarray}
where the radial part \cite{Dehesa2001} is given by
{\small \begin{eqnarray}
	\label{eq:shannonrad}
	S[\rho_{n_r,l}] &=& -\int_{\mathbb{R}_D} \rho_{n_r,l}(r)\log \rho_{n_r,l}(r)\, dr\nonumber\\ 
	&=& -\int_{0}^{\infty}2\,\omega^{D/2}\tilde{r}^{1-D/2}\omega_{l+D/2-1}(\tilde{r})[\mathcal{\tilde{L}}_{n_r}^{(l+D/2-1)}(\tilde{r})]^{2} \nonumber \\
	& & \times \log\{2\,\omega^{D/2}\tilde{r}^{1-D/2}\omega_{l+D/2-1}(\tilde{r})[\mathcal{\tilde{L}}_{n_r}^{(l+D/2-1)}(\tilde{r})]^{2}\}r^{D-1}\,dr\nonumber \\
	&=& -\log(2\,\omega^{D/2}) +\mathcal{E}\left(\mathcal{\tilde{L}}_{n_r}^{(l+D/2-1)}\right)  -l\int_{0}^{\infty}\omega_{l+D/2-1}(\tilde{r}) [\mathcal{\tilde{L}}_{n_r}^{(l+D/2-1)}(\tilde{r})]^{2}\log(\tilde{r})\,d\tilde{r} \nonumber \\
	& & +\int_{0}^{\infty} \tilde{r}\,\omega_{l+D/2-1}(\tilde{r}) [\mathcal{\tilde{L}}_{n_r}^{(l+D/2-1)}(\tilde{r})]^{2}\,d\tilde{r}\nonumber\\
	&=& 2n_r + l+D/2 -\,\log \,2 - l \,\psi(n_r+l+D/2) + \mathcal{E}\left( \tilde{{\cal{L}}}^{(l+D/2-1)}_{n_r} \right)- \frac{D}{2}\,\log \,\omega ,
	\end{eqnarray}
	where $\mathcal{E}\left(\mathcal{\tilde{L}}_{n}^{(\alpha)}\right)$ denotes the entropy of the orthonormal Laguerre polynomials $\mathcal{\tilde{L}}_{n}^{(\alpha)}(x)$ given by Equation (\ref{eq:polentropy}).\\
	On the other hand, the angular part \cite{Yanez1999,Dehesa2001} is
	\begin{eqnarray}\label{eq:shannonang}
	S[\rho_{l,\{\mu \}}] &=& - \int_{\Omega_{D-1}}  \rho_{l,\left\lbrace \mu \right\rbrace} \left( \Omega_{D-1} \right) \log \rho_{l,\left\lbrace \mu \right\rbrace} \left( \Omega_{D-1} \right) d\Omega_{D-1} \nonumber\\
	&=&-\int_{\Omega_{D-1}} |{\cal{Y}}_{l,\left\lbrace \mu \right\rbrace}(\Omega_{D-1})|^2 \log |{\cal{Y}}_{l,\left\lbrace \mu \right\rbrace}(\Omega_{D-1})|^2 d\Omega_{D-1} \equiv  \mathcal{E} \left[{\cal{Y}}_{l,\left\lbrace \mu \right\rbrace  } \right] \\
	&=& B_1 (l,\left\lbrace \mu \right\rbrace,D)+\sum^{D-2}_{j=1} \mathcal{E} \left( \tilde{C}^{(\alpha_j+\mu_{j+1})}_{\mu_j-\mu_{j+1}} \right),\qquad  D \geq 2,\label{S51}
	\end{eqnarray}
	where the symbol $\mathcal{E}\left(\tilde{C}^{(\lambda)}_{n}\right)$ denotes the entropy-like functional of the orthonormal Gegenbauer polynomial $\tilde{C}^{(\lambda)}_{n}(x)$ with respect to the weight function $\omega^{*}_{\lambda}(x)=(1-x^2)^{\lambda-1/2}$ given by Equation (\ref{eq:polentropy}), and the constant
	\begin{equation}\label{eq:B1constant}
	B_1 (l,\left\lbrace \mu \right\rbrace,D)= \log 2\pi -2 \sum^{D-2}_{j=1} \mu_{j+1} \left[\psi(2\alpha_j+\mu_j+\mu_{j+1} -\psi(\alpha_j+\mu_j)-\log 2- \frac{1}{2 (\alpha_j+\mu_j)}\right].
	\end{equation}
	Note that the angular entropy $\mathcal{E}[\mathcal{Y}_{l,\{\mu \}}]$ does not depend on $n_r$ and, moreover, its maximum value
	\begin{equation}\label{eq:EY00}
	\mathcal{E} \left[ {\cal{Y}}_{0,\left\lbrace 0 \right\rbrace}\right]=\log \frac{2 \pi^{D/2}}{\Gamma\left( \frac{D}{2} \right)},
	\end{equation} 
	which occurs for the $S$-wave states, i.e. when $(l,\left\lbrace \mu \right\rbrace)= (0,\{0\})$, is equal to $\log(2\pi)$ and $\log(4 \pi)$ for $D=2$ and $3$, respectively. See \cite{Dehesa2001,Buyarov2000,Vicente2007} for the analytical determination of the entropy-like functional of Gegenbauer polynomials.  \\
	Then, the combination of Equation (\ref{eq:shannontotal}) and (\ref{S51}) leads to the expression   
	\begin{eqnarray}
	S[\rho_{n_r,l,\{\mu \}}] &=& 2n_r + l+D/2 -\,\log \,2 - l \,\psi(n_r+l+D/2)  + B_1 (l,\left\lbrace \mu \right\rbrace,D)\nonumber \\
	& &	 +\, \mathcal{E}\left( \tilde{{\cal{L}}}^{(l+D/2-1)}_{n_r} \right) +\sum^{D-2}_{j=1} \mathcal{E} \left( \tilde{C}^{(\alpha_j+\mu_{j+1})}_{\mu_j-\mu_{j+1}} \right)- \frac{D}{2}\,\log \,\omega
	\end{eqnarray}
	for the total Shannon entropy of the $D$-dimensional oscillator state characterized by the hyperspherical quantum numbers $(n_r,l,\{\mu\})$ in position space. Similar operations get rise to the corresponding expression in momentum space $S[\rho_{n_r,l,\{\mu \}}]$, so that taking into account Equation (\ref{eq:densmom}) one has 
	\begin{equation} \label{posmomrel}
	S[\rho_{n_r,l,\{\mu \}}] + \frac{D}{2}\,\log \,\omega = S[\gamma_{n_r,l,\{\mu \}}] - \frac{D}{2}\,\log \,\omega .
	\end{equation}
	Unfortunately, these position and momentum Shannon quantities cannot be found because the involved entropy-like Laguerre and Gegenbauer polynomials are not yet known up until now \cite{Yanez1994,Dehesa2001,Dehesa1997,Buyarov2000,Vicente2007}.\\
	
	Then, to calculate the Shannon entropy for an arbitrary oscillator-like state it is more convenient to work \textbf{in Cartesian coordinates} \cite{Toranzo2019} so that, according to Equation (\ref{HOPD}), the expression (Equation (\ref{eq:posShannon})) can be written in terms of the Cartesian quantum numbers ${\{n_i\}} \equiv (n_1, n_2, \ldots,n_D)$ as
	\begin{eqnarray}
	\label{cchose1}
	S[\rho_{\{n_i\}}] &=& - \int_{\mathbb{R}_{D}} \rho_{\{n_i\}}(x_{1},\ldots,x_{D})\log [\rho_{\{n_i\}}(x_{1},\ldots,x_{D})] \, dx_{1}\cdots dx_{D} 
	= \sum_{i=1}^{3} \,\,\mathcal{I}_{D,i}
	\end{eqnarray}
	where the symbols $\{\mathcal{I}_{D,i}; i = 1,2,3\}$ denote three integral functionals of univariate Hermite polynomials. The first two integrals can be evaluated in a straightforward manner by using the elegant algebraic properties of the orthogonal Hermite polynomials \cite{Olver2010}, obtaining the expression
	\begin{align}
	\label{I11}
	\mathcal{I}_{D,1} &= -\mathcal{N}^{2}\log\left(\mathcal{N}^{2}\right)\int_{\mathbb{R}^{D}} e^{-\alpha'(x_{1}^{2}+x_{2}^{2}+\ldots+x_{D}^{2})}\,\Bigg[\Pi_{i=1}^{D}H_{n_{i}}^{2}(\sqrt{\alpha'}\, x_{i})\Bigg]\, dx_{1}\cdots dx_{D} \nonumber \\
	&=  \left(\sum_{i=1}^{D}n_{i}\right)\log 2 + \sum_{i=1}^{D} \log (n_{i}!) + \frac{D}{2} \log \left(\frac{\pi}{\alpha'}\right)
	\end{align}
	(where $\alpha' = \omega^{\frac{1}{4}}$) for the first integral, and
	\begin{align}
	\label{I21}
	\mathcal{I}_{D,2} &=\alpha'\,\mathcal{N}^{2}\int_{\mathbb{R}^{D}} e^{-\alpha'(x_{1}^{2}+x_{2}^{2}+\ldots+x_{D}^{2})}\,\Bigg[\Pi_{i=1}^{D}H_{n_{i}}^{2}(\sqrt{\alpha'}\, x_{i})\Bigg]\,(x_{1}^{2}+x_{2}^{2}+\ldots+x_{D}^{2})\, dx_{1}\ldots dx_{D} \nonumber \\
	&=  N +\frac{D}{2}
	\end{align}
	for the second integral, where $N=n_{1} +  \ldots + n_{D}$. The third integral can be expressed as
	\begin{align}
	\mathcal{I}_{D,3} &= -\mathcal{N}^{2}\int_{\mathbb{R}^{D}} e^{-\alpha'(x_{1}^{2}+x_{2}^{2}+\ldots+x_{D}^{2})}\,\Bigg[\Pi_{i=1}^{D}H_{n_{i}}^{2}(\sqrt{\alpha'}\, x_{i})\Bigg]\,\log\,\Bigg[\Pi_{i=1}^{D}H_{n_{i}}^{2}(\sqrt{\alpha'}\, x_{i})\Bigg]\, dx_{1}\cdots dx_{D}\nonumber\\
	&= -\frac{1}{\sqrt{\pi}}\Bigg[\Pi_{i=1}^{D}\frac{1}{2^{n_{i}}n_{i}!} \mathcal{E}(H_{n_i})\Bigg]
	\end{align}
	in terms of the entropy-like functional $\mathcal{E}(H_{n_i})$ of the orthogonal Hermite polynomials previously found by various authors \cite{Wolfram,Sanchez1997},
	\begin{equation}
	\mathcal{E}(H_n) \equiv \int_{0}^{\infty} [H_{n}(x)]^{2}\log[H_{n}(x)]^{2}e^{-x^{2}}\, dx = 2^{n}n!\sqrt{\pi}\log(2^{2n})-2\sum_{k=1}^{n}\,V_{n}(x_{n,k}).
	\end{equation}
	The symbol $\{x_{n,k}; k=1,\ldots,n\}$ denote the roots of the Hermite polynomial $H_n(x)$, and $\mathcal{V}_{n}(x)$ is known as the logarithmic potential of the Hermite polynomial $H_n(x)$ which is given by
	\begin{equation}\label{PLH}
	\mathcal{V}_{n}(x)=2^{n}n!\sqrt{\pi}\left[\log 2+\frac{\gamma_E}{2}-x^{2}\, {}_2 F_{2}\left(1,1;\frac{3}{2},2;-x^{2} \right)+\frac{1}{2}\sum_{i=1}^{n}\binom{n}{k}\frac{(-1)^{k}2^{k}}{k}\, {}_1 F_{1}\left(1;\frac{1}{2};-x^{2} \right) \right],	\end{equation}
	where $\gamma_E = 0.57721566$ is the Euler constant, $_1F_1(a_1;b_1;z)$ and $_2F_2(a_1,a_2;b_1,b_2;z)$ are well-known generalized hypergeometric functions \cite{Olver2010}.\\
	
	Then, by combining the Equation (\ref{cchose1}) - (\ref{PLH}) it turns out that the position Shannon entropy of the $D$-dimensional oscillator-like state is given by
	\begin{equation} \label{shapos}
	S[\rho_{\{n_{i} \}}] = A(D;\{n_i\};\{x_{n_i,i}\})\,+ \frac{D}{2} \log \left(\frac{e\,\pi}{\alpha'}\right),
	\end{equation}
	where the symbol $A(D;\{n_i\};\{x_{n_i,i}\})$ denotes
	\begin{eqnarray}\label{eseho}
	A(D;\{n_i\};\{x_{n_i,i}\}) 
	& &= N\log (2e^{1+\gamma_E}) +\sum_{i=1}^{D} \log (n_{i}!) \nonumber\\
	&  & \hspace{-1.5cm} - 2\left( \sum_{i=1}^{n_{1}}x_{n_{1},i}^{2}\, {}_2 F_{2}\left(1,1;\frac{3}{2},2;-x_{n_{1},i}^{2} \right) + \ldots + \sum_{i=1}^{n_{D}}x_{n_{D},i}^{2}\, {}_2 F_{2}\left(1,1;\frac{3}{2},2;-x_{n_{D},i}^{2} \right) \right) \nonumber \\
	& &\hspace{-3.5cm} +\sum_{k=1}^{n_{1}} {n_{1}\choose k} \frac{(-1)^{k}2^{k}}{k}\sum_{i=1}^{n_{1}}\, {}_1 F_{1}\left(k;\frac{1}{2};-x_{n_{1},i}^{2} \right)  + \ldots   +\sum_{k=1}^{n_{D}} {n_{D}\choose k} \frac{(-1)^{k}2^{k}}{k}\sum_{i=1}^{n_{D}}\, {}_1 F_{1}\left(k;\frac{1}{2};-x_{n_{D},i}^{2} \right),\nonumber\\ 
	\end{eqnarray}
	in terms of the Cartesian quantum numbers ${\{n_i\}} \equiv (n_1, n_2,\ldots,n_D)$ of the state, where $N=n_{1}+\ldots+n_{D}$. In the particular case $D=1$ one obtains
	\begin{align}
	S[\rho_{n_1}] &= \log(2^{n_1}n_1!\sqrt{\pi})+n_1+\frac{1}{2}+n_1\gamma_E-2 \sum_{i=1}^{n_1}x_{n_1,i}^{2}\, {}_2 F_{2}\left(1,1;\frac{3}{2},2;-x_{n_1,i}^{2} \right) \nonumber \\ 
	&+ \sum_{k=1}^{n_1}\binom{n_1}{k}\frac{(-1)^{k}2^{k}}{k}\sum_{i=1}^{n_1}\, {}_1 F_{1}\left(k;\frac{1}{2};-x_{n_1,i}^{2} \right),
	\end{align}
	which is the position Shannon entropy for the one-dimensional harmonic oscillator previously found \cite{Sanchez1997}.\\
	Now, taking into account the simple relation in Equation (\ref{eq:densmom}) between position and momentum oscillator densities, we can obtain with similar operations that the momentum Shannon entropy of the $D$-dimensional oscillator-like state can be expressed as
	\begin{equation} \label{shamom}
	S[\gamma_{\{n_{i} \}}] = A(D;\{n_i\};\{x_{n_i,i}\})\,+ \frac{D}{2} \log (e\,\pi\,\alpha').
	\end{equation}
	From Equation (\ref{shapos}) and (\ref{shamom}) we observe that both position and momentum Shannon entropies of the multidimensional harmonic oscillator depend only on the space dimensionality, $D$, the $D$ Cartesian quantum numbers and the location of the roots of the Hermite polynomials which control the position wavefunctions of the harmonic state. In addition, for the ground state ${\{n_i= 0; i = 1,..., D\}}$ one has that $A = 0$ because all the involved sums vanish so that we find the known values 
	\begin{equation}\label{URsumground}
	S[\rho_{\{0,\ldots,0\}}] = \frac{D}{2} \log \left(\frac{e\,\pi}{\alpha'}\right); \quad S[\gamma_{\{0,\ldots,0\}}] = \frac{D}{2} \log (e\,\pi\,\alpha')
	\end{equation}
	for the position and momentum Shannon entropies, respectively. Note that, by summing Equation (\ref{shapos}) and (\ref{shamom}), one finds the position-momentum entropic sum for the $D$-dimensional oscillator
	\begin{equation}
	S[\rho_{\{n_{i} \}}] + S[\gamma_{\{n_{i} \}}] = 2\,A(D;\{n_i\};\{x_{n_i,i}\})\,+ D\log (e\pi), 
	\end{equation}
	which fulfills the well-known Shannon-information-based uncertainty relation of Bialynicki-Birula and Mycielski (BBM) \cite{BBM1975,Beckner1995} given by 
	\begin{equation} \label{eq:BBM}
	S[\rho_{\{n_{i} \}}] + S[\gamma_{\{n_{i} \}}] \geq D\log (e\pi).
	\end{equation}
	Moreover, for the ground state ${\{n_i= 0, i = 1,..., D\}}$ one has from Equation (\ref{URsumground}) that the ground state value
	$S[\rho_{\{0,\ldots,0\}}] + S[\gamma_{\{0,\ldots,0\}}] = D \log (e\,\pi)$  saturates the BBM entropic uncertainty indeed.

	\subsection{The R\'enyi entropies}
	\label{subsec:9}

	The R\'enyi entropy for the generic $D$-dimensional oscillator-like state with the position probability density $\rho(\mathbf{r})$ given by Equation (\ref{HOPD}) and (\ref{eq:densposhyper}) in Cartesian and hyperspherical units, respectively, is defined \cite{Renyi1961} by
	\begin{equation}
	\label{eq:renentrop}
	R_{q}[\rho] =  \frac{1}{1-q}\log \int_{\mathbb{R}_D} [\rho(\mathbf{r})]^{q}\, d\mathbf{r}, \quad 0<q<\infty, \,\, q \neq 1.
	\end{equation}
	These measures quantify numerous facets of the spreading of the quantum probability density $\rho(\mathbf{r})$, which include the intrinsic randomness (uncertainty) and the geometrical profile of the quantum system. They include numerous relevant quantities as special cases, such as e.g. the disequilibrium $\langle\rho\rangle =\exp(- R_{2}[\rho])$ and the (previously considered) Shannon entropy $S[\rho] = \lim_{p\rightarrow 1} R_{p}[\rho]$. The corresponding quantities for the probability density $\gamma(\mathbf{p})$ in momentum space will be denoted by $R_{q}[\gamma]$. As measures of uncertainty the R\'enyi entropies, which have very important physico-mathematical properties \textit{per se} \cite{Aczel1975,Leonenko2008,Bialynicki2012,Dehesa2012,Jizba2004,Jizba2015}, allow for a much wider quantitative range of applicability than the Heisenberg-like measures which are based on the variance of the density and their generalizations, the radial expectation values. This allows, for example, a quantitative discussion of quantum uncertainty further beyond the conventional Heisenberg-like uncertainty \cite{Bialynicki2006,Zozor2008}. \\
	
	The computation of the R\'enyi entropies for an arbitrary oscillator-like state \textbf{in hyperspherical coordinates} have not yet been explicitly obtained by means of the state's hyperquantum numbers for reasons similar to the Shannon case. Indeed, in this case the position R\'enyi entropies for an arbitrary $D$-dimensional oscillator-like state characterized by the probability density $\rho_{n_r,l,\{\mu \}}(\mathbf{r})$ is given, according to Equation (\ref{eq:densposhyper}) and (\ref{eq:renentrop}), by
	\begin{eqnarray}
	\label{eq:posRenyi}
	R_q[\rho_{n_r,l,\{\mu \}}] &=& \frac{1}{1-q}\log \int_{\mathbb{R}_D} [\rho_{n_r,l,\{\mu \}}(\mathbf{r})]^{q}\, d\mathbf{r}\\
	\label{eq:Renyitotal}
	&=& R_q[\rho_{n_r,l}] + R_q[\rho_{l,\{\mu \}}],
	\end{eqnarray}
	where the symbols $R_q[\rho_{n_r,l}]$ and $R_q[\rho_{l,\{\mu \}}]$ denote the radial and angular R\'enyi entropies for the $D$-dimensional harmonic state, respectively. The radial  R\'enyi entropies are given \cite{Puertas2017} by
	\begin{eqnarray} \label{eq:radentropy}
	R_q[\rho_{n_r,l}] &=& \frac{1}{1-q}\log \int\limits_{0}^{\infty}[\rho_{n_r,l}(r)]^{q}\,r^{D-1}dr\nonumber\\
	&=& -\log(2\,\omega^\frac D2)+\frac1{1-q}\log N_{n_r,l}(D,q),
	\end{eqnarray}
	(we used Equation (\ref{eq:densposhyper}) to write the second equality) in which the symbol $N_{n_r,l}(D,q)$ denotes the following weighted $\mathfrak{L}_{q}$-norm of the orthogonal and orthonormal Laguerre polynomials
	\begin{eqnarray}\label{eq:radnormar}
	N_{n_r,l}(D,q) &=& \left(\frac{n_r!}{\Gamma(\alpha+n_r+1)}\right)^q\int_0^\infty r^{\alpha+lq-l}e^{-qr}\left[{\mathcal{L}}_{n_r}^{(\alpha)}(r)\right]^{2q}\,dr\nonumber\\
	&=&\int\limits_{0}^{\infty}\left(\left[\tilde{\mathcal{L}}_{n_r}^{(\alpha)}(x)\right]^{2}\,w_{\alpha}(x)\right)^{q}\,x^{\beta}\,dx\;,\quad q >0\,,
	\end{eqnarray}
	respectively, with $\alpha=l+\frac D2-1\,,\;l=0,1,2,\ldots,\, q>0\,\, \text{and}\,\, \beta=(1-q)(\alpha-l)=(p-1)(1-D/2)$.
	These values guarantee the convergence of this integral functional; i.e., the condition $\beta+q\alpha= \frac D2+lq-1 > -1$ is always satisfied for physically meaningful values of the parameters. Note that the radial  R\'enyi entropies depend on the oscillator parameters $(\omega, D)$ and the radial and orbital hyperquantum numbers $(n_r, l)$ only. \\
	The angular R\'enyi entropies are given by
	\begin{equation}
	\label{eq:renyihyd3}
	R_q[\rho_{l,\{\mu \}}] = R_{q}[\mathcal{Y}_{l,\{\mu\}}] := \frac{1}{1-q}\log \Lambda_q[\mathcal{Y}_{l,\{\mu\}}],
	\end{equation}
	where $\Lambda_q[\mathcal{Y}_{l,\{\mu\}}]$ denotes the entropic moments of the hyperspherical harmonics \cite{Puertas2017jmp},
	\begin{eqnarray}
	\label{eq:reha1}
	\Lambda_q[\mathcal{Y}_{l,\{\mu\}}] &=& \int_{\Omega_{D-1}}|\mathcal{Y}_{l,\{\mu\}}(\Omega_{D-1})|^{2q}\, d\Omega_{D-1}\nonumber\\
	&=& \mathcal{N}_{l,\{\mu \}}^{2q} \int_{\Omega_{D-1}} \prod_{j=1}^{D-2}[C^{(\alpha_{j}+\mu_{j+1})}_{\mu_{j}-\mu_{j+1}}(\cos \theta_{j})]^{2q}(\sin\theta_{j})^{2q\mu_{j+1}}\, d\Omega_{D-1}\nonumber \\
	&=& 2\pi\mathcal{N}_{l,\{\mu \}}^{2q}\prod_{j=1}^{D-2}\int_{0}^{\pi}[C^{(\alpha_{j}+\mu_{j+1})}_{\mu_{j}-\mu_{j+1}}(\cos \theta_{j})]^{2q}(\sin\theta_{j})^{2q\mu_{j+1}+2\alpha_j}\, d\theta_j,
	\end{eqnarray}
	where the normalization constant $\mathcal{N}_{l,\{\mu \}}$ is given by 
	\begin{equation}
	\label{eq:normhypersphar}
	\mathcal{N}_{l,\{\mu\}}^{2} = \frac{1}{2\pi}
	\prod_{j=1}^{D-2} \frac{(\alpha_{j}+\mu_{j})(\mu_{j}-\mu_{j+1})![\Gamma(\alpha_{j}+\mu_{j+1})]^{2}}{\pi \, 2^{1-2\alpha_{j}-2\mu_{j+1}}\Gamma(2\alpha_{j}+\mu_{j}+\mu_{j+1})}.
	\end{equation} 
	Note that, according to Equation (\ref{armonico}) and (\ref{eq:renyihyd3}), the hyperspherical harmonics $\mathcal{Y}_{l,\{\mu\}}(\Omega_{D-1})$ and consequently the angular R\'enyi entropies $R_{q}[\mathcal{Y}_{l,\{\mu\}}]$ do not depend on the radial hyperquantum number $n_r$, but they do depend on the dimensionality $D$ and the angular hyperquantum numbers.\\
	
	Similar operations in momentum space allow us to obtain the corresponding expressions for the momentum R\'enyi entropies $R_q[\gamma_{n_r,l,\{\mu \}}]$ for any $D$-dimensional harmonic state, so that taking into account Equation (\ref{eq:densmom}) one has 
	\begin{equation} \label{eq:posmomREL}
	R_q[\rho_{n_r,l,\{\mu \}}] + \frac{D}{2}\,\log \,\omega = R_q[\gamma_{n_r,l,\{\mu \}}] - \frac{D}{2}\,\log \,\omega .
	\end{equation}
	From Equation (\ref{eq:radentropy}) and (\ref{eq:reha1}) one can note that the determination of the radial and angular R\'enyi entropies of the $D$-dimensional harmonic systems in hyperspherical coordinates are controlled  by some power-like integral functionals of the Laguerre and Gegenbauer polynomials, which have not yet explicitly  found. Let us here point out that they can be analytically tackled by two recent methodologies: one based on the Srivastava's linearization method \cite{Sanchez2013} and another one based on the combinatorial Bell polynomials \cite{Sanchez2010}. To illustrate that this is possible, we tackle in subsection \ref{subsec:10} the calculation of the disequilibrium of the $D$-dimensional harmonic systems  which corresponds to the simplest case $q=2$ of the R\'enyi entropies.\\
	
	Then, it seems natural to compute the R\'enyi entropies for an arbitrary oscillator-like state by operating \textbf{in Cartesian units} \cite{Puertas2018}. So, let us now calculate the R\'enyi entropies $R_{q}[\rho_{\{n_i\}}], q \neq 1,$ given by Equation (\ref{eq:renentrop}) for a generic state characterized by the Cartesian quantum numbers ${\{n_i\}} \equiv (n_1, n_2,\ldots,n_D)$. This quantity, according to Equation (\ref{HOPD}),  can be expressed as
	\begin{align}
	\label{HORE}
	R_{q}[\rho_{\{n_i\}}] &= \frac{1}{1-q}\log \int_{-\infty}^{\infty} dx_{1}\ldots \int_{-\infty}^{\infty} dx_{D} \, [\rho_{\{n_i\}}](\mathbf{r})]^{q} \nonumber \\
	&= \frac{1}{1-q}\log\left( \mathcal{N}^{2q}\Bigg[\Pi_{i=1}^{D}\int_{-\infty}^{\infty} e^{-\alpha' q x_{i}^{2}}\left|H_{n_{i}}(\sqrt{\alpha'}\, x_{i})\right|^{2q} \, dx_{i}\Bigg]\right) .
	\end{align}
	Then, we calculate these integral functionals Hermite polynomials by linearizing the involved Hermite powers \textit{\`a la Srivastava}, which means \cite{Sanchez2013} by using the following linearization relation for the $(2q)$-th power of the Hermite polynomials 
	\begin{equation}
	\label{linforH2}
	\left|H_{n}\left(\sqrt{\alpha'} x\right)\right|^{2q} = A_{n,q}(\nu)q^{-q\nu}\sum_{j=0}^{\infty}\frac{1}{(-1)^{}2^{2j } j!}c_{j}\left(q\nu,2q,\frac{1}{q},\frac{n-\nu}{2},\nu-\frac{1}{2},-\frac{1}{2} \right) H_{2j}(\sqrt{\alpha' q}x),
	\end{equation}
	
	with the expansion coefficients
	\begin{align}\label{coeff}
	\hspace{-3cm}c_{j}&\left( q\nu,2q,\frac{1}{q},\frac{n-\nu}{2},\nu-\frac{1}{2},-\frac{1}{2} \right)= \nonumber \\
	& = \left(\frac{1}{2}\right)_{q\nu} \binom{\frac{n+\nu-1 }{2}}{\frac{n-\nu}{2}}^{2q} F_{A}^{(2q+1)}\left( \begin{array}{cc}
	q\nu+\frac{1}{2} ; \overbrace{\frac{\nu-n }{2}, \ldots, \frac{\nu - n}{2}}^{2q}, -j & \\
	&; \underbrace{\frac{1}{q}, \ldots, \frac{1}{q}}_{2q},1\\
	\underbrace{\nu + \frac{1}{2}, \ldots, \nu+\frac{1}{2}}_{2q},\frac{1}{2} & \\
	\end{array}\right),
	\end{align}
	where the symbol $F_A^{(s)}(x_1,\ldots,x_s)$ denotes the Lauricella function of type A of $s$ variables and $2s+1$ parameters defined as \cite{Sanchez2010,Srivastava1985}
	\setlength{\tabcolsep}{4pt} 
	\renewcommand{\arraystretch}{1.5} 
	\begin{equation}
	F_A^{(s)}
	\left( 
	\begin{array}{cc}
	a; b_1,\ldots,b_s\\
	c_1,\ldots,c_s
	\end{array}
	; x_1,\ldots,x_s
	\right)
	=
	\sum_{j_1,\ldots,j_s=0}^{\infty}
	\frac{(a)_{j_1+\cdots+j_s} (b_1)_{j_1} \cdots (b_s)_{j_s}}{(c_1)_{j_1}
		\cdots (c_s)_{j_s} }\frac{x_1^{j_1}\cdots x_s^{j_s}}{j_1!\cdots j_s!}
	\label{eq:lauricella_definition}
	\end{equation}
	and $(z)_a = \frac{\Gamma(z+a)}{\Gamma(z)}$ is the known Pochhammer's symbol. \\
	
	Finally, the combination of Equation (\ref{HORE}) and (\ref{coeff}) allows us to express the R\'enyi entropy of order $q$ for the oscillator-like state with Cartesian quantum numbers ${\{n_i\}} \equiv (n_1, n_2,\ldots,n_D)$ as follows
	
	\begin{eqnarray}
	\label{HORE3}\nonumber
	R_{q}[\rho_{\{n_i\}}] 
	&=&    -\frac D2 \log\left[\alpha'\right]+\mathcal K_q\,D+\overline{\mathcal K}_q\,N_O+\frac {q}{q-1}\sum_{i=1}^D(-1)^{n_i}\log\left[\left(\frac{n_i+1}{2}\right)_{\frac12}  \right]+\frac{1}{1-q}\sum_{i=1}^D\log\left[\mathfrak F_q(n_i)\right], \nonumber\\
	\end{eqnarray}
	where 
	\begin{equation}
	\mathcal K_q=\frac{\log[\pi^{q-\frac12}\,q^\frac12]}{q-1}; \quad \overline{\mathcal K}_q=\frac{1 }{1-q}\log \left[\frac{4^{q}\,\Gamma\left(\frac12+q\right)}{\pi^{\frac{1}{2}}\,q^{q}}\right]
	\end{equation}
	and the symbol $\mathfrak F_q(n_i)$ denotes the following multivariate Lauricella function of type A:
	\begin{align}
	\label{HOLF2}
	\mathfrak F_q(n)&\equiv 
	F_{A}^{(2q+1)}\left( \begin{array}{cc}
	q\nu+\frac{1}{2} ; \frac{\nu-n }{2}, \ldots, \frac{\nu - n}{2},0 & \\
	&; \frac{1}{q}, \ldots, \frac{1}{q},1\\
	\nu+ \frac{1}{2}, \ldots, \nu+\frac{1}{2},\frac{1}{2} & \\
	\end{array}\right) =
	F_{A}^{(2q)}\left( \begin{array}{cc}
	q\nu+\frac{1}{2} ; \frac{\nu-n }{2}, \ldots, \frac{\nu - n}{2} & \\
	&; \frac{1}{q}, \ldots, \frac{1}{q}\\
	\nu + \frac{1}{2}, \ldots, \nu +\frac{1}{2} & \\
	\end{array}\right) &\nonumber\\
	&=  \sum_{j_{1}, \ldots, j_{2q}=0 }^{\infty} \frac{\left(q\nu+\frac{1}{2}\right)_{j_{1}+\ldots j_{2q}} (\frac{\nu-n }{2})_{j_{1}} \cdots (\frac{\nu-n }{2})_{j_{2q}} }{(\nu + \frac{1}{2})_{j_{1}} \cdots (\nu+ \frac{1}{2})_{j_{2q}} } \frac{\left(\frac{1}{q}\right)^{j_{1}} \cdots \left(\frac{1}{q}\right)^{j_{2q}}}{j_{1}!\cdots j_{2q}! }&
	\nonumber\\
	&=  \sum_{j_{1}, \ldots, j_{2q}=0 }^{\frac{n-\nu}2} \frac{\left(q\nu+\frac{1}{2}\right)_{j_{1}+\ldots j_{2q}} (\frac{\nu-n }{2})_{j_{1}} \cdots (\frac{\nu-n }{2})_{j_{2q}} }{(\nu + \frac{1}{2})_{j_{1}} \cdots (\nu + \frac{1}{2})_{j_{2q}} } \frac{\left(\frac{1}{q}\right)^{j_{1}} \cdots \left(\frac{1}{q}\right)^{j_{2q}}}{j_{1}!\cdots j_{2q}! }.
	\end{align}
	Note that this Lauricella function is reduced to a finite sum because $\frac{\nu-n}{2}$ is always a negative integer number. Moreover, for convenience, we have used the notation $N_O=\sum_{i=1}^D\nu_i$ for the amount of odd numbers $n_i$; so that $N_E=D-N_O$ gives the number of the even ones.\\
	Summarizing, the expression in Equation (\ref{HORE3}) allows for the analytical determination of the R\'enyi entropies (with positive integer values of $q$) for any arbitrary state of the multidimensional harmonic systems. For the ground state (i.e., $n_i=0,\,i=1,\ldots, D$; so, $N=0$) this general expression boils down to  
	\begin{equation}
	R_q[\rho_{\{0\}}]=\frac D2\log\left[\frac {\pi\, q^{\frac1{q-1}}}{\alpha'}\right].
	\end{equation}
	In fact, this ground state R\'enyi entropy holds for any $q>0$ as one can directly obtain from Equation (\ref{HORE}).\\
	Taking into account that the momentum density is a re-scaled form of the position density, we have the following expression for the associated momentum R\'enyi entropy,
	\begin{eqnarray}
	\label{Remsp}
	R_{\tilde q}[\gamma_{\{n_i\}}] &=&  \frac D2 \log\left[\alpha'\right]+\mathcal K_{\tilde q}\,D+\overline{\mathcal K}_{\tilde q}\,N_O+\frac {\tilde q}{\tilde q-1}\sum_{i=1}^D(-1)^{n_i}\log\left[\left(\frac{n_i+1}{2}\right)_{\frac12}  \right]+\frac{1}{1-\tilde q}\sum_{i=1}^D\log\left[\mathfrak F_{\tilde q}(n_i)\right],\nonumber\\
	\end{eqnarray}
	($\tilde q\in \mathbb{N}$). Although Equation (\ref{HORE3}) and (\ref{Remsp}) rigorously hold for $q\not=1$ and $q\in\mathbb N$ only, it is reasonable to conjecture its general validity for any $q>0, \,q\not=1$ provided the formal existence of a generalized function $\mathfrak F_q(n)$. 
	
	Finally, from Equation (\ref{HORE3}) and (\ref{Remsp}) we have that the general expression for the position-momentum R\'enyi entropic sum is
	\begin{eqnarray}\nonumber
	R_{q}[\rho_{\{n_i\}}]+R_{\tilde q}[\gamma_{\{n_i\}}] &=&  (\mathcal K_{q}+\mathcal K_{\tilde q})\,D+(\overline{\mathcal K}_{q}+\overline{\mathcal K}_{\tilde q})\,N_O+\left(\frac {q}{ q-1}+\frac {\tilde q}{\tilde q-1}\right)\sum_{i=1}^D(-1)^{n_i}\log\left[\left(\frac{n_i+1}{2}\right)_{\frac12}  \right]
	\\
	&+&\frac{1}{1- q}\sum_{i=1}^D\log\left[\mathfrak F_{q}(n_i)\right]+\frac{1}{1-\tilde q}\sum_{i=1}^D\log\left[\mathfrak F_{\tilde q}(n_i)\right],
	\end{eqnarray}
	which verifies the R\'enyi-entropy-based uncertainty relation of Zozor-Portesi-Vignat \cite{Zozor2008} when $\frac1q+\frac1{\tilde q}\ge2$ for arbitrary quantum systems.
	In the conjugated case $\tilde q=q^*$ such that $\frac1q+\frac1{q^*}=2$, one obtains
	\begin{eqnarray}\nonumber
	R_{q}[\rho_{\{n_i\}}]+R_{q^*}[\gamma_{\{n_i\}}] &=&  D\log\left(\pi q^{\frac1{2q-2}}{q^*}^{\frac{1}{2q^*-2}}\right)+(\overline{\mathcal K}_{q}+\overline{\mathcal K}_{ q^*})\,N_O\\
	&+&\frac{1}{1- q}\sum_{i=1}^D\log\left[\mathfrak F_{q}(n_i)\right]+\frac{1}{1- q^*}\sum_{i=1}^D\log\left[\mathfrak F_{ q^*}(n_i)\right].
	\end{eqnarray}
	Let us finally remark that the first term corresponds to the sharp bound for the general R\'enyi-entropy-based uncertainty relation 
	\begin{equation}\label{eq:BialR}
	R_{q}[\rho_{\{n_i\}}]+R_{q^*}[\gamma_{\{n_i\}}] \ge D\log\left(\pi q^{\frac1{2q-2}}{q^*}^{\frac{1}{2q^*-2}}\right)
	\end{equation}
	with conjugated parameters \cite{Bialynicki2006,Zozor2008}.
	
	\subsection{The disequilibrium}
	\label{subsec:10}
	
	Let us now calculate in hyperspherical units the disequilibrium for a $D$-dimensional oscillator-like state with the position probability density $\rho(\mathbf{r}) \equiv \rho_{n_r,l,\{\mu \}}(\mathbf{r})$ given by Equation (\ref{eq:densposhyper}). This quantity (also called informational energy), which quantifies the departure of $\rho(\mathbf{r})$ from equiprobability, is defined as
	\begin{eqnarray}
	\label{eq:diseq}
	\mathcal{D}[\rho] & \equiv & \langle \rho \rangle = \int_{\mathbb{R}_D} [\rho(\mathbf{r})]^{2}\, d\mathbf{r}
	\end{eqnarray} 
	So, it is closely related to the second order R\'enyi entropy as mentioned above. From Equation (\ref{eq:densposhyper}) and (\ref{eq:diseq}), the disequilibrium can be expressed as
	\begin{equation}
	\mathcal{D}[\rho_{n_r,l,\{\mu \}}] = \mathcal{D}[\rho_{n_r,l}] \times \mathcal{D}[\rho_{l,\{\mu \}}] \end{equation}
	where
	\begin{eqnarray}
	\label{eq:harmoscdiseqrad}
	\mathcal{D}[\rho_{n_r,l}] & \equiv & \langle \rho_{n_r,l} \rangle = \left[\frac{2\, n_r! \,\omega^{l+\frac{D}{2}}}{\Gamma\left(n_r+l+\frac{D}{2}\right)}\right]^{2}\int_{0}^{\infty} r^{4l+D-1}e^{-2\omega r^{2}}  \left[L_{n_r}^{(l+D/2-1)}(\omega r^{2})\right]^{4} \,dr 
	\end{eqnarray}
	and 
	\begin{eqnarray}
	\label{eq:harmoscdiseqang}
	\mathcal{D}[\rho_{l,\{\mu \}}] &=& \langle \rho_{l,\{\mu \}} \rangle = \int_{\Omega_{D}}|\mathcal{Y}_{l,\{\mu\}}(\Omega_{D})|^{4}\, d\Omega_{D}.   
	\end{eqnarray}
	are the radial and angular parts of the disequilibrium, respectively.\\
	To evaluate the radial part $\mathcal{D}[\rho_{n_r,l}]$ we apply twice the linearization formula (see e.g. \cite{Sanchez1999} and references therein)
	\begin{eqnarray}
	\label{eq:linearformlaguer}
	[L_{n}^{(\alpha)}(x) ]^{2} &= & \frac{\Gamma(\alpha+1+n)}{2^{2n}n!}\sum_{k=0}^{n}\binom{2n-2k}{n-k}\frac{(2k)!}{k!}  \frac{1}{\Gamma(\alpha+1+k)}L_{2k}^{(2\alpha)}(x), \nonumber 
	\end{eqnarray}
	and then we use the following integral involving the product of two Laguerre polynomials (see e.g. \cite{Sanchez1999} and references therein)
	\begin{eqnarray}
	\label{eq:inttwolaguer}
	\int_{0}^{\infty}x^{s}e^{-x}L_{n}^{(\alpha)}(x)L_{m}^{(\beta)}(x)\, dx &=& \Gamma(s+1)\sum_{r=0}^{\min(2k,2k')}(-1)^{m+n}\binom{s-\alpha}{n-r}\binom{s-\beta}{m-r}\binom{s+r}{r},\nonumber 
	\end{eqnarray}
	to finally obtain the expression
	{\small \begin{eqnarray}
		\label{eq:harmoscdiseqrad2}
		\mathcal{D}[\rho_{n_r,l}] &=& \omega ^{\frac{D}{2}} 2^{1-\frac{D}{2}-l-4 n_r} \,\Gamma\left(\frac{D}{2}+2 l\right)   \sum_{k,k'=0}^{n_r}\sum_{r=0}^{\min(2k,2k')}\binom{2n_r-2k}{n_r-k}\binom{2n_r-2k'}{n_r-k'}\nonumber\\
		&  &\times\frac{(2k)!}{k!}\frac{(2k')!}{k'!}\frac{1}{\Gamma(l+D/2+k)}\frac{1}{\Gamma(l+D/2+k')}\binom{1-D/2}{2k-r}\binom{1-D/2}{2k'-r}\binom{2l+D/2-1+r}{r}.\nonumber \\
		\end{eqnarray}}
	For physical interest and checking, let us point out that for the ground state ($n_r=0,l=0$) we have that $\mathcal{D}[\rho_{0,0}] = \frac{ \omega ^{D/2}}{2^{\frac{D}{2}-1}}$.\\
	
	To evaluate the angular part $\mathcal{D}[\rho_{l,\{\mu \}}]$ given by Equation (\ref{eq:harmoscdiseqang}), we use the expression (Equation (\ref{armonico})) of the hyperspherical harmonics and the orthonormal Gegenbauer polynomials together with the $D$-dimensional solid angle element to write that\\
	\begin{align}\nonumber
	\mathcal{D}[\rho_{l,\{\mu \}}]&=\frac{1}{2 \pi}
	\prod^{D-2}_{j=1}\int^{+1}_{-1}
	|\tilde{C}_{\mu_j-\mu_{j+1}}^{(\alpha_j+\mu_{j+1})}(x_j)|^4
	\omega_{\alpha_j+2\mu_{j+1}}(x_j) dx_j\\ \label{average2}
	&=\frac{1}{2 \pi} \prod^{D-2}_{j=1} \int^{+1}_{-1} |\tilde{C}_{n}^{(\lambda)}(x_j)|^4 \omega_{\lambda+\mu_{j+1}}(x_j) dx_j,
	\end{align}
	with the change of variable $\theta_j\rightarrow x_j=\cos \theta_j$ in the first equality, and the notation $\lambda=\alpha_j+2\mu_{j+1}$ and $n=\mu_j-\mu_{j+1}$ in the second equality. To evaluate the integral involved in Equation (\ref{average2}) we shall use twice the following Dougall's linearization formula for the orthonormal Gegenbauer polynomials \cite{Dehesa2007jmp}
	\begin{equation}\label{transCorto}
	\left[\tilde{C}_n^{(\lambda)}(x) \right]^2=\sum^{n}_{k=0} b(\lambda, \lambda+\mu_{j+1}, n; k) \tilde{C}_{2k}^{(\lambda+\mu_{j+1})}(x),
	\end{equation}
	where
	\begin{align} \nonumber
	b(\lambda, \lambda+\mu_{j+1}, n; k) & = \frac{(n+\lambda) \Gamma(k+1/2) \Gamma(k+\lambda) \Gamma(k+n+2\lambda) \Gamma(\lambda+\mu_{j+1})}{\pi^{1/2} \Gamma(1-k+n)\Gamma(k+\lambda+1/2)\Gamma(k+2 \lambda)\Gamma(2k+\lambda+\mu_{j+1})}\\\nonumber
	&\qquad \times \left( \frac{2^{1-2\lambda-2\mu_{j+1}} \Gamma(2k+2\lambda+2\mu_{j+1})}{(2k+\lambda+\mu_{j+1})\Gamma(2k+1) \Gamma^2(\lambda+\mu_{j+1})}\right)^{1/2}  \\\label{coefOrto}
	&\qquad \quad \times {}_4F_3 \left( \begin{array}{c|}
	k-n,\quad k+n+2\lambda,\quad k+\lambda,\quad k+\lambda+\mu_{j+1}+\frac{1}{2}\quad \\
	2k+\lambda+\mu_{j+1}+1,\quad k+2\lambda,\quad k+\lambda+\frac{1}{2}
	\end{array} \ 1\right),
	\end{align}
	and the symbol $_4F_3(1)$ denotes the well-known generalized hypergeometric function $_4F_3(z)$ \cite{Olver2010} evaluated at $z=1$.
	Then, we finally obtain the expression \cite{Dehesa2007jmp}
	\begin{align}\nonumber
	\mathcal{D}[\rho_{l,\{\mu \}}] & =\frac{1}{2 \pi} \prod^{D-2}_{j=1} \sum^{n}_{k=0} b^2(\lambda, \lambda+\mu_{j+1}, n; k)\\\label{averageFinal}
	& =\frac{1}{2 \pi} \prod^{D-2}_{j=1} \sum^{\mu_j-\mu_{j+1}}_{k=0} b^2(\alpha_j+\mu_{j+1}, \alpha_j+2\mu_{j+1}, \mu_j-\mu_{j+1}; k)
	\end{align}
	for the angular part of the disequilibrium of any $D$-dimensional stationary state of a central potential. This expression can be simplified much more because the function $_4F_3(z)$ boils down to a $_3F_2(z)$ as indicated in \cite{Dehesa2007jmp}. Even more, the resulting expression can be much further reduced for specific states with $D=2$ and $3$, obtaining $\mathcal{D}[\rho_{l}]  =\frac{1}{2 \pi}$, and 
	\begin{equation}
	\label{eq:harmoscdiseqang2}
	\mathcal{D}[\rho_{l,m}] = \sum_{l'=0}^{2l}
	\left(\frac{\hat{l}^{2}\hat{l}'}{\sqrt{4\pi}}
	\right)^{2}\left(\begin{array}{ccc}
	l & l & l' \\
	0 & 0& 0			\end{array}\right)^{2}
	\left(\begin{array}{ccc}
	l & l & l' \\
	m & m& -2m			\end{array}\right)^{2},
	\end{equation}
	respectively, where the notation $\hat{l} = \sqrt{2l+1}$ and the $3j$-symbols \cite{Olver2010} have been used. In fact the last expression can be alternatively derived by using the integrals of the product of four hyperspherical harmonics given by Equation (\ref{eq:harmoscdiseqang}), and similar integrals with a bigger number of hyperspherical harmonics, which appear in classical and quantum-mechanical  non-relativistic \cite{Avery2006} and relativistic \cite{Coelho2002,Kyriakopoulos1968} problems. Note that for $S$-wave states ($l=m=0$) we have that $\mathcal{D}[\rho_{0,0}] = 0$.

	\section{The entropy-like measures of Rydberg oscillator states}
	\label{sec:4}
	
	The dispersion facets of the multidimensional spreading for the Rydberg oscillator-like states (i.e., states with high and very high radial hyperspherical quantum number $n_r$) have been previously determined in subsection \ref{subsec:4} by means of the radial expectation values. Later, in subsection \ref{subsec:5}, we can observe that the multidimensional Rydberg spreading facets given by the (local) Fisher informations have the values $F_{\rho}^{(R)}  \sim 4n_r \, \omega$ and $F_{\gamma}^{(R)}  \sim \frac{4n_r}{\omega}$ in position and momentum spaces, respectively.  In this section we will study the information-theoretical facets of the multidimensional Rydberg spreading by means of the (global) Shannon and R\'enyi entropies of the quantum probability densities given by Equation (\ref{eq:densposhyper}) and (\ref{eq:densmom}).
	
	\subsection{The Shannon entropy of Rydberg states} 
	\label{subsec:11}
	
	The Shannon entropy for an arbitrary $D$-dimensional oscillator-like state characterized by the hyperspherical quantum numbers $(n_r,l,\left\lbrace \mu \right\rbrace)\equiv(n_r,\mu_1,\mu_2,\ldots,\mu_{D-1})$  is given, according to Equation (\ref{eq:shannontotal}) and (\ref{eq:shannonang}), by
	\begin{eqnarray}
	\label{eq:posShannonto}
	S[\rho_{n_r,l,\{\mu \}}] + \frac{D}{2}\,\log \,\omega &=& 2n_r + l+D/2 -\,\log \,2 - l \,\psi(n_r+l+D/2) + \mathcal{E}\left( \tilde{{\cal{L}}}^{(l+D/2-1)}_{n_r} \right) +  \mathcal{E}[\mathcal{Y}_{l,\{\mu \}}].\nonumber\\
	\end{eqnarray}
	To evaluate this expression for Rydberg states (i.e., in the limit $n_r \rightarrow \infty$) we take into account the asymptotics of the digamma or Psi function $\psi\left(l+\frac{D}{2}+n_r \right)\sim \log n_r$ \cite{Olver2010} and the strong asymptotics of Laguerre polynomials \cite{Aptekarev1995,Dehesa1998} which gives the following asymptotic behavior for the entropy-like functional  $\mathcal{E}_{\beta}\left(\mathcal{\tilde{L}}_{n}^{(\alpha)}\right)$  of orthonormal Laguerre polynomials \cite{Dehesa1998}:  
	\begin{eqnarray}
	\label{eq:asympgenshan}
	\mathcal{E}_{\beta}\left(\mathcal{\tilde{L}}_{n}^{(\alpha)}\right) &:=& \int_{0}^{\infty} x^{\beta}\omega_{\alpha}(x)[\mathcal{\tilde{L}}_{n}^{(\alpha)}(x)]^{2}\log\,[\mathcal{\tilde{L}}_{n}^{(\alpha)}(x)]^{2}\, dx\nonumber \\
	&=& \frac{2^{2\beta+2}\Gamma(\beta+3/2)}{\sqrt{\pi}\,\Gamma(\beta+2)}n^{\beta+1}-\frac{2^{2\beta}(\alpha+1)\Gamma(\beta+1/2)}{\sqrt{\pi}\,\Gamma(\beta+1)}n^{\beta}\log n + \frac{2^{2\beta-1}\Gamma(\beta+1/2)}{\sqrt{\pi}\,\Gamma(\beta+1)}[2(\alpha+1)\psi(\beta+1)\nonumber\\
	& &-(2\alpha+1)\psi(\beta+1/2)-2\log\pi -4(\alpha-1)\log 2  + \gamma_{E}+4+2(\alpha+2\beta)+4\alpha\beta]n^{\beta}+o(n^{\beta}),\nonumber \\
	\end{eqnarray}
	which holds for any real $\alpha>-1$. From this expression, since $\mathcal{E}_{\beta}\left(\mathcal{\tilde{L}}_{n}^{(\alpha)}\right):=-\mathcal{E}_{0}\left(\mathcal{\tilde{L}}_{n}^{(\alpha)}\right)$, we have the following asymptotical behavior for the entropy of orthonormal Laguerre polynomials 
	\begin{equation}
	\label{eq:asympentrolag}
	\mathcal{E}\left(\mathcal{\tilde{L}}_{n_r}^{(\alpha)}\right) = -2n_r+(\alpha+1)\log n_r - \alpha -2 +\log(2\pi) + o(1), 
	\end{equation}
	so that from Equation (\ref{eq:posShannonos}) and (\ref{eq:asympentrolag}) we finally have the expression 
	\begin{equation} \label{eq:posentrRy}
	S[\rho_{n_r,l,\{\mu \}}] + \frac{D}{2}\,\log \,\omega \simeq 	\frac{D}{2}\log n_r +\log\pi -1 +  \mathcal{E}[\mathcal{Y}_{l,\{\mu \}}]
	\end{equation}
	for the dominant term of the position Shannon entropy for the $D$-dimensional oscillator-like state of Rydberg type, where the angular part $\mathcal{E}[\mathcal{Y}_{l,\{\mu \}}]$ is under control. Moreover, working similarly in momentum space, one has 
	\begin{equation}
	S[\gamma_{n_r,l,\{\mu \}}] - \frac{D}{2}\,\log \,\omega \simeq 	\frac{D}{2}\log n_r +\log\pi -1 +  \mathcal{E}[\mathcal{Y}_{l,\{\mu \}}]
	\end{equation}
	for the dominant term of the momentum Shannon entropy for the $D$-dimensional oscillator-like state of Rydberg type, in consistence with the previous general expression in Equation (\ref{posmomrel}). The sum of these two last expressions finally gives the position-momentum Shannon-information sum for all stationary Rydberg states $(n_r>>1,l,\left\lbrace \mu \right\rbrace)$ of oscillators with a given dimenionality $D$:
	\begin{equation}
	S[\rho_{n_r,l,\{\mu \}}] + S[\gamma_{n_r,l,\{\mu \}}] \simeq \log n_r + 2\left(\log\pi -1 +  \mathcal{E}[\mathcal{Y}_{l,\{\mu \}}]\right),
	\end{equation}
	where the angular Shannon entropy $\mathcal{E}[\mathcal{Y}_{l,\{\mu \}}]$, which depend on $(l,\left\lbrace \mu \right\rbrace),D)$ but not on $n_r$, is given by Equation (\ref{eq:shannonang}) and (\ref{eq:EY00}). Finally, let us highlight that this sum fulfills not only the Shannon-entropy uncertainty relation in Equation (\ref{eq:BBM}) which holds for all $D$-dimensional quantum systems, but also the tighter Shannon-entropy uncertainty relation 
	\begin{equation}
	S[\rho_{n_r,l,\{\mu \}}] + S[\gamma_{n_r,l,\{\mu \}}] \ge C_{l,\{\mu\}}
	\label{eq:shannon_uncertainty_relation}
	\end{equation}
	where
	\begin{eqnarray}
	C_{l,\{\mu\}}&=& 2l+D+2\log\left[\frac{\Gamma\left(l+\frac{D}{2}\right)}{2}\right]-(2l+D-1)\psi\left(l+\frac{D}{2}\right)\nonumber\\
	&&+(D-1)\left(\psi\left(\frac{2l+D}{4}\right)
	+\log 2\right) + 2 \, \mathcal{E}[\mathcal{Y}_{l,\{\mu \}}],
	\label{eq:central_bound}
	\end{eqnarray}
	valid for all $D$- dimensional quantum systems subject to arbitrary central potentials \cite{Rudnicki2012}.

	\subsection{The R\'enyi entropy of Rydberg states}
	\label{subsec:12}
	
	
	The R\'enyi entropies for an arbitrary $D$-dimensional oscillator-like state characterized by the hyperspherical quantum numbers $(n_r,l,\left\lbrace \mu \right\rbrace)\equiv(n_r,\mu_1,\mu_2,\ldots,\mu_{D-1})$ are given, according to Equation (\ref{eq:Renyitotal}), (\ref{eq:radentropy}) and (\ref{eq:renyihyd3}), by
	\begin{equation}
	R_{q}[\rho_{n_r,l,\{\mu \}}] = R_{q}[\rho_{n_r,l}]+R_{q}[\mathcal{Y}_{l,\{\mu \}}] = -\log(2\,\omega^\frac D2)+\frac1{1-q}\log N_{n_r,l}(D,q) + R_{q}[\mathcal{Y}_{l,\{\mu \}}].
	\end{equation}
	To evaluate this expression for Rydberg states we have to determine the asymptotics of the weighted $\mathfrak{L}_{q}$-norm of the orthonormal Laguerre polynomials, $N_{n_r,l}(D,q)$, given by Equation (\ref{eq:radnormar}), in the limit $n_r \rightarrow \infty$. This has been  recently done \cite{Aptekarev2016} by using the theory of the strong asymptotics of Laguerre polynomials \cite{Aptekarev1995,Dehesa1998} obtaining that the R\'enyi entropies of an arbitrary $D$-dimensional harmonic state are
	\begin{equation} \label{eq:totalRenyiRyd}
	R_{q}[\rho_{n_r,l,\{\mu \}}] + \frac{D}{2}\log\omega \simeq \frac1{1-q}\log N_{asymp}(n_r,l,D,q) + R_{q}[\mathcal{Y}_{l,\{\mu \}}],
	\end{equation}
	where the symbol $N_{asymp}(n_r,l,D,q)$ denotes the asymptotical value of the weighted Laguerre norm $N_{n_r,l}(D,q)$ in the limit $n_r \rightarrow \infty$, which is given by
	\begin{equation}\label{8}
	\resizebox{0.8\textwidth}{!}{$N_{asymp}(n_r,l,D,q) \left\{
		\begin{array}{ll}
		\sim C(\beta,q)\,(2n_r)^{(1-q)\,D/2}\,,\quad & q\in(0,q^{*})\\
		\displaystyle=\frac{2}{\pi^{q+1/2}n_r^{q/2}}\,\displaystyle\frac{\Gamma(q+1/2)}{\Gamma(q+1)}\,(\log n_r+ O(1))\,,\quad & q=q^{*}\\
		\sim C_{B}(\alpha,\beta,q)\,n_r^{(q-1)D/2-q}\,,\quad & q>q^{*}
		\end{array}
		\right.\;$}
	\end{equation}
	for $q>0, n_r >>1, \,l=0,1,2,\ldots$ and $D > 2$ \cite{Aptekarev2016}. The symbols $q^{*}:=\frac{D}{D-1}$, $\alpha=l+\frac D2-1,\, \beta=(1-q)(\alpha-l)=(q-1)(1-D/2)$, and the constants $C$ and $C_{B}$  are given by  
	\begin{equation}\label{4}
	C_{B}(\alpha,\beta,q):=2\int\limits_{0}^{\infty}t^{2\beta+1}|J_{\alpha} (2t)|^{2q}\,dt;\quad C(\beta,q):=\displaystyle\frac{2^{\beta+1}}{\pi^{q+1/2}}\,\displaystyle\frac{\Gamma(\beta+1-q/2)\,\Gamma(1-q/2)\,\Gamma(q+1/2)}
	{\Gamma(\beta+2-q)\,\Gamma(1+q)}\; ,
	\end{equation}
	where $J_{\alpha}(z)$ denotes the Bessel function \cite{Olver2010}.  Note that $N_{asymp}(n_r,l,D,q)$ is constant (i.e., independent of $n_r$) and equal to $C_{B}(\alpha,\beta,q)$ only when $(q-1)D/2-q = 0$. This means that the constancy occurs either when $D=\frac{2q}{q-1}$, or $q=\frac{D}{D-2}$. Moreover the angular R\'enyi entropies, $R_{q}[\mathcal{Y}_{l,\{\mu \}}]$, which are given by Equation (\ref{eq:renyihyd3}) and (\ref{eq:reha1}), do not depend on the radial hyperquanyum number $n_r$. They can be analytically  expressed in terms of the angular hyperquantum numbers by means of the  linearization of the powers of the involved Gegenbauer polynomials either in terms of some generalized hypergeometric functions of Srivastava-Karlsson type \cite{Sanchez2013} or via the combinatorial Bell polynomials \cite{Dehesa2017}. In addition, the R\'enyi entropies for the Rydberg states of three-dimensional ($D=3$) isotropic harmonic oscillator are discussed monographically in \cite{Dehesa2017}, where the angular part $R_{q}[\mathcal{Y}_{l,m}]$ is also explicitly given in terms of the orbital and magnetic quantum numbers by the two different methods just mentioned. \\
	
	For the remaining cases $D = 2$ and $D\in[0,2)$, the total R\'enyi entropies of any $D$-dimensional oscillator-like state of Rydberg type (i.e., characterized with the hyperspherical quantum numbers $(n_r>>1,l,\left\lbrace \mu \right\rbrace)$ are given by Equation (\ref{eq:totalRenyiRyd}), where the asymptotical value $N_{asymp}(n_r,l,D,q)$ is explicitly given in refs. \cite{Aptekarev2016,Aptekarev2012}, respectively. The R\'enyi entropies for the Rydberg states of one-dimensional ($D=1$) isotropic harmonic oscillator (1D-HO) are examined monographically in \cite{Aptekarev2012} by means of the strong asymptotics of the weighted $\mathfrak{L}_{q}$-norm of the Hermite polynomials, because these polynomials control the wavefunctions of all the stationary states of the 1D-HO. \\
	
	Taking into account Equation (\ref{eq:posmomREL}), one has that the momentum R\'enyi entropies for the Rydberg states is obtained from the position R\'enyi entropies as
	\begin{equation}
	\label{OMRE1}
	R_{q}[\gamma_{n_r,l}] = R_{q}[\rho_{n_r,l}] + D\log \omega,
	\end{equation}
	so that the position-momentum R\'enyi-entropy sum for the Rydberg harmonic states is
	\begin{equation}
	\label{OREUS1}	
	R_{q}[\rho_{n_r,l,\{\mu \}}] + R_{p}[\gamma_{n_r,l,\{\mu \}}] \simeq  \frac2{1-q}\log N_{asymp}(n_r,l,D,q) +2\,R_{q}[\mathcal{Y}_{l,\{\mu \}}], 
	\end{equation}
	which holds for $q>0, n_r >>1, \,l=0,1,2,\ldots$ and the asymptotical values $N_{asymp}(n_r,l,D,q)$ have been given above. As expected, note that this sum does not depend on the oscillator strength $\omega$. Moreover, it can be shown that it fulfills not only the general R\'enyi entropy uncertainty relation for multidimensional quantum systems \cite{Bialynicki2006,Zozor2008}
	\begin{equation}\nonumber
	R_{q_1}[\rho_{\{n_i\}}]+R_{q_2}[\gamma_{\{n_i\}}] \ge D\log\left(\pi q_1^{\frac1{2q_1-2}}{q_2}^{\frac{1}{2q_2-2}}\right),
	\end{equation}
	with the conjugated parameters $q_1$ and $q_2$, but also the (conjectured) R\'enyi entropy uncertainty relation for multidimensional quantum systems subject to a central potential \cite{Dehesa2020}.
	
	
	\section{Dispersion and information-theoretical properties of high-dimensional oscillator states}
	\label{sec:5}
	
	The dispersion facets of the multidimensional spreading for the pseudo-classical oscillator-like states (i.e., states with high and very high $D$) have been previously determined in subsection \ref{subsec:5} and \ref{subsec:6} by means of the radial expectation values and their Heisenberg uncertainty product. Later, in subsection \ref{subsec:7}, we have observed that the high-dimensional spreading facets given by the (local) Fisher informations have the values $F_{\rho}^{(R)}  \sim 2\,D \, \omega$ and $F_{\gamma}^{(R)}  \sim \frac{2D}{\omega}$ in position and momentum spaces, respectively.  In this section we will study the information-theoretical facets of the high-dimensional spreading by means of the (global) Shannon and R\'enyi entropies of the quantum probability densities given by Equation (\ref{eq:densposhyper}) and (\ref{eq:densmom}).

	\subsection{The Shannon entropy of high-dimensional oscillators}
	\label{subsec:13}
	

	Here let us show that the position Shannon entropy of the stationary $D$-dimensional oscillator-like states with given hyperquantum numbers $(n_r,l,\left\lbrace \mu \right\rbrace)$ in the limit $D\rightarrow \infty$ is given by
	\begin{equation}\label{eq:infty_label}
	S_{n_r,l,\{\mu\}}(D) + \frac{D}{2}\log\,\omega\,=\, \frac{1}{2}D\, \log D \,+ \,O(D).
	\end{equation}
	To obtain this result we start from the general expression given in Equation (\ref{eq:posShannonto}) for the position Shannon entropy  of the $D$-dimensional harmonic  state $(n_r,l,\left\lbrace \mu \right\rbrace)$,
	\begin{eqnarray}
	S[\rho_{n_r,l,\{\mu \}}] + \frac{D}{2}\,\log \,\omega &=& 2n_r + l+D/2 -\,\log \,2 - l \,\psi(n_r+l+D/2) + \mathcal{E}\left( \tilde{{\cal{L}}}^{(l+D/2-1)}_{n_r} \right) +  \mathcal{E}[\mathcal{Y}_{l,\{\mu \}}]\nonumber\\
	&=& A_{2}(n_r,l,D) + B (l,\left\lbrace \mu \right\rbrace,D) + E\left( \tilde{{\cal{L}}}^{(\alpha)}_{n_r} \right) + \sum_{j=1}^{D-2} E\left[  \tilde{{\cal{C}}}_{\mu_j-\mu_{j+1}}^{(\alpha_j+\mu_{j+1})}\right]
	\label{eq:Shannonto2}
	\end{eqnarray}
	with $\alpha = l+\frac{D}{2}-1$, $2\alpha_{j}=D-j-1$, $A_{2}(n_r,l,D)=2n_r + \alpha +1 -\log\, 2 - l \psi(n_r+\alpha+1) $ and the constant $B (l,\left\lbrace \mu \right\rbrace,D)$ is given by Equation (\ref{eq:B1constant}). Now, we have to evaluate the four terms of Equation (\ref{eq:Shannonto2}) in the limit $D\rightarrow \infty$, obtaining \cite{Dehesa2019} the expression
	\begin{eqnarray}\label{eq:Sro2har}
	S[\rho_{n_r,l,\{\mu \}}] + \frac{D}{2}\log \,\omega &\sim& A_{2,\infty}+ B _{\infty} +E\left( \tilde{{\cal{L}}}_{\infty} \right) + E(\widetilde{C}_{\infty}),
	\end{eqnarray}
	with the following values
	\begin{eqnarray} 
	A_{2,\infty} =\lim_{D\to\infty} A_2 (n_r,l,D) &=& \frac{D}{2}-l \log \left(\frac{D}{2}\right) -l (n_r+l-1/2)\frac{2}{D}+ \log \left(\frac{e^{2n_r+l}}{2} \right)=\frac{D}2+o(D), \label{Ainf}\\
	B_{\infty} &=&\lim_{D\to\infty} B (l,\left\lbrace \mu \right\rbrace,D) = 2 \sum^{D-2}_{j=1} n_r
	\left[\,\log\alpha_j\, -\, \log \alpha_j\, +\,O(1)\,\right] = \, O(D)
	\end{eqnarray}
	for the first two terms (which follow from the asymptotics of the digamma function mentioned above), and 
	\begin{eqnarray} 
	E(\tilde{\cal{L}}_{\infty})  &=&\lim_{D\to\infty} E\left( \tilde{{\cal{L}}}^{(\alpha)}_{n_r-l-1} \right)\Big|_{\alpha=l+D/2-1} = \,\frac12 D \log D - \frac{\log 2 + 1}2 D + O (\log D) \label{eq:asyLaguerre}\\ 
	E(\tilde{\cal{C}}_{\infty}) &=& \lim_{D\to\infty}\sum^{D-2}_{j=1} E \left( {\tilde{\cal{C}}}^{(\alpha_j+\mu_{j+1})}_{\mu_j-\mu_{j+1}} \right) \, =\, O (\log D),\label{eq:asyGegen}
	\end{eqnarray}
	for the remaining two terms which correspond to the asymptotical values of the orthonormal Laguerre polynomials and the summation of $(D-1)$ orthonormal Gegenbauer polynomials, respectively.\label{eq:asyGegenbauer} \\
	
	To obtain the expression given by Equation (\ref{eq:asyLaguerre}) we have used the parameter-asymptotics \cite{Dehesa2019} for the Shannon-like integral functional of the orthogonal Laguerre polynomials $\mathcal{L}^{(\alpha)}_{m}(x), m\in\mathbb{Z_{+}}$, which gives
	\begin{eqnarray}
	\label{eq:shalaguerreasy}
	I(\mathcal{L}^{(\alpha)}_{m},\sigma) &=& \int_{0}^{\infty} x^{\alpha+\sigma-1}e^{- x}\left[\mathcal{L}^{(\alpha)}_{m}(x)\right]^{2}\log\left[\mathcal{L}^{(\alpha)}_{m}(x)\right]^{2}\, dx, \nonumber\\
	&\sim& \frac{\sqrt{2\pi}}{(m-1)!}\, \left(\frac{\alpha}{e}\right)^{\alpha}\, \alpha^{\sigma + m-\frac{1}{2}} \log\,\alpha,  \quad \alpha\rightarrow\infty,
	\label{eq:finalresult}
	\end{eqnarray}
	with $\sigma\in\mathbb{R}$, where the symbol $A \sim B$ means that $A/B \to 1$. Then, according to Equation (\ref{eq:polentropy}) we have
	\begin{eqnarray}
	E\left( \tilde{{\cal{L}}}^{(\alpha)}_{n_r} \right)
	&= &\frac{n_r!E\left( {\cal{L}}^{(\alpha)}_{n_r} \right)}{\Gamma(n_r+\alpha+1)}
	- \log\left[\frac{n_r!}{\Gamma(n_r+\alpha+1)}\right]
	\nonumber \\
	&=& - \frac{n_r!\Gamma(\alpha+1)\alpha^{n_r}\log\alpha}{(n-1)!\Gamma(n_r+\alpha+1)}
	- \log\left[\frac{n_r!}{\Gamma(n_r+\alpha+1)}\right]
	\nonumber \\
	&\sim& \log(\Gamma(n_r+\alpha+1)) - \log n_r! - n_r \log \alpha
	\nonumber \\
	&=& \alpha\log\alpha - \alpha + \frac 12 \log\alpha
	+O(1)\nonumber \\
	& =& \frac{1}2 D \log D\,-\, \frac{\log 2 +1}2D\,+\,\frac12 \log D\,+\,O(1), \nonumber
	\label{eq:El}
	\end{eqnarray}
	where we used $\alpha:=D/2+l-1$, $E\left( {\cal{L}}^{(\alpha)}_{n} \right)= -I(\mathcal{L}^{(\alpha)}_{n},1)$, and the asymptotics~$\log\Gamma(z)=\big(z-\frac 12\big)\log z - z + O(1)$ as~$z\to\infty$.
	
	To prove the remaining expression given by Equation (\ref{eq:asyGegen}) we start from the Shannon-like integral functional of the orthonormal Gegenbauer polynomials
	\begin{equation}
	\label{eq:shanp2}
	E\left(\tilde{C}^{(\alpha)}_{m}\right) = -\int_{-1}^{1} (1-x^{2})^{\alpha-\frac{1}{2}}[\tilde{C}^{(\alpha)}_{m}(x)]^{2}\log \left[\tilde{C}^{(\alpha)}_{m}(x)\right]^{2},
	\end{equation}
	which has the following parameter-asymptotical behavior \cite{Dehesa2019}
	\begin{eqnarray}
	\label{eq:shanp22}
	E\left(\tilde{C}^{(\alpha)}_{m}\right) &\sim& -2 \Bigg[\log(m+2\alpha-1,m)   +\frac{m}{2}\Big(\psi\left(m+\frac{1}{2}\right)-\psi(m+\alpha+1)\Big) \Bigg]\nonumber \\
	&  & \sim -m\log\alpha + \frac{m(m+3)}{2\alpha} - \frac{m}{2(\alpha+m+1)} -2m\log 2 -m\psi\left(m+\frac{1}{2}\right) +\log[\Gamma(m+1)] \nonumber \\
	& & \sim -m\log\alpha + \frac{m(m+2)}{2\alpha}  -2m\log 2-m\psi\left(m+\frac{1}{2}\right) +\log[\Gamma(m+1)], \quad \alpha\rightarrow\infty.
	\end{eqnarray}
	Returning to the asymptotics of the angular Shannon entropy we notice that the sum $\sum_{j=1}^{D-2} \mathcal{E}\left[ \tilde{C}_{\mu_j-\mu_{j+1}}^{(\alpha_j+\mu_{j+1})}\right]$ from Equation (\ref{eq:Shannonto2}) consists at most of $n$ non-zero terms. It is because of $\mathcal{E}\left[\tilde{C}_{0}^{(\alpha_j)}\right]=0$ and
	$\,\,l\equiv\mu_1\geq\mu_2\geq\ldots\geq \left|\mu_{D-1} \right|\geq 0$. Thus, taking into account the notation  $2\alpha_{j} = D-j-1$, and the corollary from Equation (\ref{eq:shanp22})
	\[
	\mathcal{E}(\tilde{C}^{(\alpha)}_{m})\,=\,O(\log \alpha),
	\]
	we obtain the wanted $(D\rightarrow\infty)$-asymptotical behavior given by Equation (\ref{eq:asyGegen}).\\
	
	Finally, in momentum space one can operate similarly to obtain
	\begin{equation} \label{eq:asyparamom}
	S[\gamma_{n_r,l,\{\mu \}}] - \frac{D}{2}\,\log \,\omega = \frac{1}{2}D\, \log D \,+ \,O(D), 
	\end{equation}
	for the dominant term of the momentum Shannon entropy for the $D$-dimensional oscillator-like state with high and very high $D$, in consistence with the previous general expression given by Equation (\ref{posmomrel}). The sum of the position and momentum expressions in Equation (\ref{eq:infty_label}) and (\ref{eq:asyparamom}) finally gives the following \textbf{position-momentum Shannon-information sum for all stationary high and very high dimensionality states, i.e.,$D\rightarrow \infty$ and fixed hyperquantum numbers $(n_r>>1,l,\left\lbrace \mu \right\rbrace)$}:
	\begin{equation}
	S[\rho_{n_r,l,\{\mu \}}] + S[\gamma_{n_r,l,\{\mu \}}] = D\, \log D \,+ \,O(D).
	\end{equation}
	Note that this sum fulfills not only the Shannon-entropy uncertainty relation given in Equation (\ref{eq:BBM}), which holds for all $D$-dimensional quantum systems, but also the tighter Shannon-entropy uncertainty relation \cite{Rudnicki2012} given by Equation (\ref{eq:shannon_uncertainty_relation}) and (\ref{eq:central_bound}).

	\subsection{The R\'enyi entropy of high-dimensional oscillators}
	\label{subsec:14}
	
	
	
	The R\'enyi entropies for an arbitrary $D$-dimensional oscillator-like state characterized by the hyperspherical quantum numbers $(n_r,l,\left\lbrace \mu \right\rbrace)$ are given, according to Equation  (\ref{eq:Renyitotal}) and (\ref{eq:renyihyd3}), by
	\begin{equation}
	\label{eq:renyihyd1}
	R_{q}[\rho_{n_r,l,\{\mu\}}] = R_{q}[\rho_{n_r,l}]+R_{q}[\mathcal{Y}_{l,\{\mu\}}].
	\end{equation}
	The radial part $R_{q}[\rho_{n_r,l}]$ can be expressed  according to Equation (\ref{eq:radentropy}), as
	\begin{equation}
	\label{eq:renyihyd4}
	R_{q}[\rho_{n_r,l}] = -\log(2\,\omega^\frac D2)+\frac1{1-q}\log N_{n_r,l}(D,q),
	\end{equation}
	where $N_{n_r,l}(D,q)$ denotes the following weighted $\mathfrak{L}_{q}$-norm of the orthogonal Laguerre polynomials:
	\begin{eqnarray}\label{eq:radnormaoL}
	N_{n_r,l}(D,q) &=& \left(\frac{n_r!}{\Gamma(\alpha+n_r+1)}\right)^q\int_0^\infty r^{\alpha+lq-l}e^{-qr}\left[{\mathcal{L}}_{n_r}^{(\alpha)}(r)\right]^{2q}\,dr,\quad q >0\,,
	\end{eqnarray}
	with $\alpha=l+\frac D2-1\,,\;l=0,1,2,\ldots,\, q>0\,\, \text{and}\,\, \beta=(1-q)(\alpha-l)=(p-1)(1-D/2)$. The angular part, $R_{q}[\mathcal{Y}_{l,\{\mu \}}]$, is given by Equation (\ref{eq:renyihyd3}) and (\ref{eq:reha1}) in terms of some weighted $\mathfrak{L}_{q}$-norm of the orthogonal Gegenbauer polynomials. \\
	
	Here we will determine the $(D\rightarrow\infty)$-asymptotical behavior of the position and momentum R\'enyi entropies of $D$-dimensional oscillator-like systems.  In position space this problem reduces, according to Equation (\ref{eq:renyihyd1}), to the evaluation of the radial and angular R\'enyi entropies in the asymptotical limit $(D\rightarrow\infty)$. This is done by using the parameter-asymptotics of the $\mathfrak{L}_{q}$-norms of Laguerre and Gegenbauer polynomials recently developed \cite{Temme2017,Puertas2017}. First, we observe that the  $\mathfrak{L}_{q}$-norm of the orthogonal Laguerre polynomials behaves asymptotically as
	\begin{equation}
	\label{eq:ren6}
	N_{n_r,l}(D,q) \sim \frac{\sqrt{2\pi}}{(n_r!)^q}\,q^{l(1-q)-1}\,\left(\frac{|q-1|}{q}\right)^{2qn}\,\frac{\alpha^{\alpha+q(l+2n_r)-l+\frac12}}{[\Gamma(\alpha+ n_r+1)]^q}(qe)^{-\alpha},
	\end{equation}
	so that, according to Equation (\ref{eq:renyihyd4}), one has that the radial entropy behaves as 
	\begin{eqnarray}
	\label{eq:ren9bis}
	R_q[\rho_{n_r,l}]&\sim& \frac{1}{2} D \log D + \frac{1}{2} \log \left(\frac{q^{\frac1{q-1}}}{2\,\omega \,e}\right)\, D +\left(\frac{q\,n_r}{1-q}-\frac12\right)\log D,
	\end{eqnarray} 
	which holds for $q>0, q\neq 1$. Moreover, similar operations with the parameter-asymptotics of the $\mathfrak{L}_{q}$-norms of Gegenbauer polynomials allow for the following asymptotical behavior,
	\begin{eqnarray}\label{eq:Rq_ang}
	R_{q}[\mathcal{Y}_{l,\{\mu\}}] 
	&\sim &-\frac{1}{2}D\log D +\frac{1}{2}D\log(2e\pi) +\frac{1}{2}\log D  +\frac 1{1-q}\log\left(\tilde{\mathcal E}(D,\{\mu\})^q\tilde{\mathcal M}(D,q,\{\mu\})\right),\nonumber \\
	\end{eqnarray}
	of the angular R\'enyi entropy of the generic harmonic state with angular hyperquantum numbers $(l,\{\mu\})$, which holds for every non-negative $q \neq 1$. In Equation (\ref{eq:Rq_ang}) we have used the following constants
	\begin{eqnarray}\label{eq:Etilde}
	\tilde{\mathcal E}(D,\{\mu\})
	&=&\prod_{j=1}^{D-2}(\alpha_j+\mu_{j+1})^{2(\mu_j-\mu_{j+1})} \frac{\Gamma(2\alpha_j+2\mu_{j+1})}{\Gamma(2\alpha_j+\mu_{j+1}+\mu_j)}\frac{\Gamma(\alpha_j+\mu_{j+1})}{\Gamma(\alpha_j+\mu_{j})}
	\end{eqnarray}
	and
	\begin{equation}\label{eq:Mtilde}
	\tilde{\mathcal M}(D,q,\{\mu\})\equiv 4^{q(l-\mu_{D-1})} \pi^{1-\frac D2}\prod_{j=1}^{D-2}\frac{\Gamma\big(q(\mu_j-\mu_{j+1})+\frac12\big)}{\Gamma\big(\mu_j-\mu_{j+1}+1\big)^q}.
	\end{equation}
	Note that $\tilde{\mathcal E}=\tilde{\mathcal M}=1$ for any configuration with $\mu_1=\mu_2=\cdots=\mu_{D-1}$.  Moreover, we have the values
	\begin{eqnarray}
	R_{q}[\mathcal{Y}_{0,\{0\}}]&\sim & \log \left(\frac{2\pi^\frac D2}{\Gamma\left(\frac D2\right)}\right)\nonumber\\
	&\sim& -\frac D2\log\left(\frac D2\right)+\frac D2 \log (e\pi )+\frac12\log\left(\frac D2\right)\nonumber\\
	&\sim &-\frac{1}{2}D\log D +\frac{1}{2}D\log(2e\pi) +\frac{1}{2}\log D
	\label{eq:reha5}
	\end{eqnarray}
	and 
	\begin{eqnarray}
	\label{eq:reha6}
	R_{q}[\mathcal{Y}_{n-1,\{n-1\}}]  
	&\sim & -\frac{1}{2}D\log D +\frac{1}{2}D\log(2e\pi) +\frac{1}{2}\log D +\frac{1}{1-q}\log \left(\frac{\Gamma ((n-1) q+1)}{\Gamma (n)^q}\right)
	\end{eqnarray}
	for the $(ns)$ and circular states of high-dimensional harmonic states, respectively. See \cite{Puertas2017} for further details and to know how to determine more terms in these asymptotical radial and angular developments.\\
	
	Then, putting the radial and angular asymptotical expressions in Equation (\ref{eq:ren9bis}) and (\ref{eq:Rq_ang}) into Equation (\ref{eq:renyihyd1}) we obtain the following value for the total position R\'enyi entropies for high-dimensional harmonic states:
	\begin{align}
	\label{ORE1}
	R_{q}[\rho_{n_r,l,\{\mu \}}] + \frac{D}{2}\log\omega &\sim \frac{D}{2}\log\left(q^{\frac{1}{q-1}}\pi \right)+\frac{q\,n_r}{1-q}\log D\nonumber\\
	& + \frac{1}{1-q}\log(\tilde{\mathcal{E}}(D,\{\mu \})^{q}\tilde{M}(D,q,\{\mu \} )\,\hat{\mathfrak{C}}(n_r,l,q)\,2^{-q\,n_r})  
	\end{align}
	where the constants, $\tilde{\mathcal{E}}(D,\{\mu \})$ and $\tilde{M}(D,q,\{\mu \} )$, are given by Equation (\ref{eq:Mtilde}) and (\ref{eq:Etilde}), respectively, and $\hat{\mathfrak{C}}(n_r,l,q) = \frac{2^{q-1}}{(n_r!)^{q}} q^{-q (2 \,n_r+l)}\left| q-1\right| ^{2 \,n_r\, q} $. Further algebraic manipulations have allowed for the following closed compact expression
	\begin{align}\label{eq:TotSpaGS}
	R_{q}[\rho_{n_r,l,\{\mu \}}] + \frac{D}{2}\log\omega &\simeq \frac{D}{2}\log\left(q^{\frac{1}{q-1}}\pi \right) + \mathcal{O}(\log\,D),\quad q\not=1
	\end{align}
	for the R\'enyi entropies of any high-dimensional oscillator-like state with fixed hyperquantum numbers in position space. Taking into account Equation (\ref{eq:posmomREL}), one has that the momentum R\'enyi entropies for the high-dimensional harmonic states have the expression
	\begin{align}
	\label{ORE2}
	R_q[\gamma_{n_r,l,\{\mu\}}]- \frac{D}{2}\log\omega \sim \frac D2\log\left(q^{\frac1{q-1}}\pi\right)+\frac{q\,n_r}{1-q}\log D, \quad q\not=1
	\end{align}
	for the R\'enyi entropies of high-dimensional oscillator-like states in momentum space. Note that, in both position and momentum R\'enyi entropies, the oscillator-strength  appears early in the dominant term and the dependence on the quantum numbers $(n_r,l)$ does not appear up to the second term. In addition, we can also observe from Equation (\ref{ORE1}) and (\ref{ORE2}) that in the limit $q\rightarrow 1$ one has 
	\begin{align}
	\label{OSE1}
	S[\rho_{n_r,l,\{\mu \}}] &\sim \frac D2\log\left(\frac{e\pi}{\omega}\right), \\
	\label{OSE2}
	S[\gamma_{n_r,l,\{\mu \}}] &\sim \frac D2\log\left(e\pi\omega\right)
	\end{align}
	for the position and momentum Shannon entropy of a general $(n_r,l,\{\mu\})$-state of the high-dimensional harmonic system.
	
	The summation of the expressions Equation (\ref{ORE1}) and (\ref{ORE2}) allows us to have the following leading term for the joint position-momentum R\'enyi uncertainty sum of a high-dimensional harmonic system \cite{Sobrino2017} for a general state $(n_r,l,\{\mu \})$: 
	\begin{equation}
	\label{renyiHO}
	R_p[\rho_{n_r,l,\{\mu\}}]+R_q[\gamma_{n_r,l\{\mu\}}] \sim D\log\left(\pi p^{\frac1{2(p-1)}}q^{\frac1{2(q-1)}}\right); \qquad D\rightarrow\infty,
	\end{equation}
	where $\frac 1p+\frac1q=2$ (i.e., $\frac {p}{p-1}+\frac{q}{q-1}=0$ or $q=\frac {p}{2p-1}$), which saturates the  R\'enyi-entropy-based uncertainty relation \cite{Zozor2008} 
	\begin{eqnarray}
	\label{GREUP}
	R_{q}[\rho]+R_{p}[\gamma] \geq D\log\left(p^{\frac{1}{2(p-1)}}q^{\frac{1}{2(q-1)}}\pi\right),
	\end{eqnarray}
	that holds for general quantum systems. Moreover, note that in the limiting case with $q\to1$ and $p\to1$, the expression Equation (\ref{renyiHO}) gives the following dominant term of the position-momentum Shannon-entropy-based uncertainty sum  
	\begin{equation}
	\label{eq:USS}
	S[\rho_{n_r,l,\{\mu \}}] + S[\gamma_{n_r,l,\{\mu \}}]  \sim  D(1+\log\pi)
	\end{equation}
	which saturates the position-momentum Shannon-uncertainty-based relation in Equation (\ref{eq:BBM})  \cite{BBM1975}, what is a further checking of our results.

\section{Conclusion}
\label{conclusions}

We describe an analytical approach to determine the dispersion and entropy-like measures of multidimensional harmonic oscillator systems directly in terms of the hyperquantum numbers of the stationary states, the space dimensionality and the potential strength. These physical measures quantify the various spreading/delocalization facets of the harmonic states, which are characterized by their quantum probability densities in position and momentum spaces. This approach is based on the algebraic and asymptotical properties of the special functions of mathematical physics (hyperspherical harmonics, classical orthogonal polynomials, generalized hypergeometric functions) which are involved in the oscillator states' wavefunctions.\\

For arbitrary oscillator states, whose wavefunctions are controlled by the Laguerre and Gegenbauer polynomials in hyperspherical units and by Hermite polynomials in Cartesian units, we conclude that the radial expectation values and the Fisher information are best calculated in hyperspherical units, see Equations (\ref{eq:defhf0}) and (\ref{eq:defhf}), as well as Equations (\ref{eq:fishinf1}) and (\ref{eq:fishinf2}). This is basically because of the gradient-like functional form of the kinetic-energy operator and the Fisher information. However, the Shannon and R\'enyi entropies of positive integer order and the disequilibrium for these states cannot be calculated in these units because they are controlled by some logarithmic-like and power-like integral functionals of Laguerre and Gegenbauer polynomials (then, ultimately by weighted $\mathcal{L}_q$-norms of these orthogonal polynomials) whose determination is a mathematical open problem nowadays despite so many efforts.  On the contrary, in Cartesian units the Shannon entropy can be calculated, see Equations (\ref{shapos}) and (\ref{shamom}), by means of the well-known roots of the Hermite polynomials with degrees equal to the state's hyperquantum numbers. Moreover, by using the linearization of powers of Hermite polynomials, the R\'enyi entropies of positive integer order can be found by Equations (\ref{HORE3}) and (\ref{Remsp}) in terms of the Cartesian hyperquantum numbers. In addition, the associated Heisenberg-like and entropic uncertainty relations have been examined and discussed.\\

For highly-excited Rydberg states we have shown that the dispersion and information-theoretical measures can be expressed in a closed compact form by means of the hyperspherical quantum numbers. Indeed, they are given by Equation (\ref{eq:asympr1}) for the radial expectation values thanks to the weak* degree-asymptotics of some power-like functionals Laguerre polynomials, and by Equations (\ref{eq:posentrRy}) and (\ref{eq:totalRenyiRyd}) for the R\'enyi and Shannon measures with the help of the strong degree-asymptotics of some entropy-like functionals of Laguerre polynomials.\\

For high-dimensional oscillator systems we have explicitly determined the dominant term of the radial expectation values and the Shannon and R\'enyi entropies  in terms of the  hyperquantum numbers by means of Equations (\ref{7}), (\ref{eq:infty_label}) and  (\ref{eq:TotSpaGS}), respectively, by means of the parameter-asymptotics of some power- and entropy-like  functionals of Laguerre polynomials. Moreover, indications to obtain further asymptotical terms are also given.\\

We emphasize that this approach is not intended to replace proper, linear-algebra-based treatments of the localization/delocalization topic. Instead it is meant to offer a brief, rigorous yet accessible and hands-on approach to it for multidimensional harmonic oscillators, that will hopefully motivate other researchers to pursue in the future more in-depth localization/delocalization studies for other relevant quantum systems whose Schr\"{o}dinger equations are analytically solvable by means of the special functions of mathematical physics.\\

Finally, a number of issues remain open. For example, the determination of the Shannon and R\'enyi entropies directly by means of the hyperspherical quantum numbers have not yet been found for \textit{arbitrary} harmonic oscillator states up until now. Basically, the reason is that these quantities in hyperspherical units are naturally expressed in terms of the logarithmic and power-like integral functionals of the Laguerre and Gegenbauer orthogonal polynomials whose analytical evaluation is not yet explicitly known. We have illustrated this difficult task by analyzing in detail the evaluation of the simplest second-order R\'enyi entropy, the disequilibrium, in both hypersphherical and Cartesian coordinate systems. Moreover, it would be very helpful \textit{per se} and for its applications (a) to quantify the internal complexity of the multidimensional quantum systems, and (b) to extend the present information-theoretically-based spreading study to reference modelling systems other than the harmonic oscillator. This is analytically feasible at least for the small bunch of \textit{elementary} multidimensional quantum potentials which are used to approximate the mean-field potential of the physical many-body systems, such as e.g. the potentials of zero-range, Coulomb, pseudo-harmonic, Kratzer, Morse,... (see e.g., \cite{Dong2011,Aquino2010,Oyewumi2008,Buyukasik2016,Adegoke2016,Yamano2018,Talukdar2020,Aljaber2008,Demiralp2005,Patil2007,Najafizade2016}); then, the associated Schr\"odinger equation is analytically solvable so that the quantum states of these reference systems are described by wavefunctions which are controlled by special functions of mathematical physics. As well, it would be interesting to extend the generalized Heisenberg expressions here found to quantum information \cite{Mandilara2012}. Finally, the extension of the stationary study here shown and reviewed remain to be done for the time-dependent multidimensional harmonic oscillators as well as to Dirac oscillators and some anharmonic oscillators \cite{Farina2020,Nascimento2018,Nascimento2017,Montanez2020,Dong2005}.

\section*{Acknowledgements} 
This work has been partially supported by the Grant FIS2017-89349P of the Agencia Estatal de Investigaci\'on (Spain)) and the European Regional Development Fund (FEDER), and the Grant FQM-207 of the Agencia de Innovaci\'on y Desarrollo  de Andaluc\'ia .


%
%
%

%
%

\end{document}